# Application of boundary functionals of random processes in statistical physics

V. V. Ryazanov

Institute for Nuclear Research, pr. Nauki, 47 Kiev, Ukraine, e-mail: vryazan19@gmail.com

The potential applications of boundary functionals of random processes, such as the extreme values of these processes, the moment of first reaching a fixed level, the value of the process at the moment of reaching the level, the moment of reaching extreme values, the time the process stays above a fixed level, and other functionals, are considered for the description of physical, chemical, and biological problems. Definitions of these functionals are provided, and characteristic functions are presented for the model with an exponential distribution of incoming demands. A generalization of these limitations is also considered. The potential uses of boundary functionals are demonstrated through examples such as a unicyclic network with affinity $A$, an asymmetric random walk, nonlinear diffusion, two-level model, Brownian motion, and multiple diffusing particles with reversible target-binding kinetics.

## 1. Introduction

Random walks and processes with independent increments on finite Markov chains, as well as semi-Markov processes described by sums of random variables, generalize classical homogeneous processes with independent increments and are used to model queueing theory, risk theory, stochastic storage theory, and reliability theory. These processes are also widely used to solve various physical, chemical, and biological problems [1-8].

Boundary value problems for these processes are of considerable theoretical and practical interest, as they involve studying the distributions of functionals such as extreme values of the processes, the moment of first reaching a fixed level, the value of the process at the moment of reaching that level, the moment of reaching extreme values, the time the process spends above a fixed level, and other related functionals. For example, key characteristics studied in risk theory are closely related to the distributions of boundary functionals for homogeneous processes with independent increments, such as extrema over finite intervals, absolute extrema, jump functionals, and others. Different aspects of the study of boundary functionals, using various mathematical approaches, are described in many works (see, e.g., Refs. [1-21, 66, 67].

The most extensively studied boundary functional is the first-passage time (*FPT*) [22-58], which is used across various fields such as physics, chemistry, biology, and economics. The volume of work on *FPT* research exceeds several thousand papers. The list of existing and emerging areas of *FPT* research is also extensive. Many articles and books are also dedicated to the study of extreme values of random functions and processes. However, most of these applications pertain to mathematical statistics and related fields, while applications to physical or biological problems are relatively fewer, though still more numerous compared to other boundary functionals. Other boundary functionals are less frequently used compared to *FPT*, although their importance is arguably not less significant. In works [59-65], *FPT* was incorporated into statistical distributions as a thermodynamic parameter. It is possible to replace *FPT* with other boundary functionals in these distributions and apply them to study the behavior of a physical quantity chosen within the statistical distribution. For example, one could examine the effect of entropy change on the chosen functional, as was done for *FPT* in papers [60-62, 64-65]. A promising area of research could be the combined application of several boundary functionals to a given phenomenon.

In this article, we build on risk theory, whose primary application lies in the financial sector. However, risk theory is closely related to other fields such as renewal theory and applied mathematical disciplines like queueing theory, stochastic storage theory, and reliability theory. These disciplines often solve their problems by referring to risk theory, and conversely, risk theory draws on general mathematical principles and finds applications in these areas. Risk theory, as its name suggests, examines phenomena such as the presence of a risk process in survival and risk zones. This type of behavior is relevant to many physical, chemical, and biological processes and phenomena.

In this article, several boundary functionals are examined, with definitions provided and their behavior illustrated in relation to the conjugate thermodynamic parameter, which acts as a thermodynamic force. Their physical significance is explained. Some boundary functionals apparently have not been applied to physical problems before. Sections 4 and 5 are specifically devoted to examples of their potential applications in physical problems. The use of these functionals in risk theory, economics, queueing theory, and other applied mathematical disciplines has long been successfully practiced. Extreme value statistics are also applied in a wide range of fields, including physics.

The paper is organized as follows. Section 2 provides definitions of several boundary functionals. Section 3 presents the characteristic functions (*CF*) and moment generating functions of these functionals for the exponential distribution model of incoming demands. Section 4 explores the application of these functionals to a unicyclic network with affinity *A* and to an asymmetric random walk. Section 5 applies the sojourn time functional above a given level to nonlinear diffusion and to multiple diffusing particle with reversible target-binding kinetics. Other tasks have been considered, particularly the two-level model and Brownian motion. The approximations made are discussed. Finally, Section 6 offers brief concluding remarks.

**Notations and definitions**

The article considers random processes with independent increments $\{\xi(t),\ 0 \le t \le T\}$ defined on the probability space $\{\Omega, \mathcal{F}, \boldsymbol{P}\}$ (works [9-10]) with real values in the phase space ($R$, $\mathcal{B}$), $R = (-\infty, \infty)$, $\mathcal{B}$ is the $\sigma$-algebra of Borel sets on the line $R$. Here $\Omega$ is the space of elementary events. For example, $\Omega = \{\omega_1, \omega_2, ...\}$, $\boldsymbol{P}(\{\omega_k\}) = p_k$ in the discrete case, $\{\omega_k\}$ is a one-point set containing $\omega_k$, $\mathcal{F}$ is the $\sigma$-algebra of subsets of $\Omega$; elements of this $\sigma$-algebra are called (random) events; $\boldsymbol{P}$ is a probability - measure on $\{\Omega, \mathcal{F}\}$ such that $\boldsymbol{P}\{\Omega\} = 1$. We assume that the process $\{\xi(t),\ 0 \le t \le T\}$ is separable, and its sample functions belong to the space $D_{[0,T]}$ of right-continuous functions without discontinuity of the second kind with finite left-side boundaries.

A process $\{\xi(t),\ 0 \le t \le T\}$ is called a process with independent increments if for an arbitrary $n$≥1 and for all $\{t_k\}_{k=\overline{0,n}}$, $0 \le t_0 < t_1 < ... < t_n \le T$, the values $\xi(t_0)$ and increments $\xi(t_0)$, $\xi(t_1)$ - $\xi(t_0)$,…, $\xi(t_n) - \xi(t_{n-1})$ are independent random variables.

A process with independent increments is called homogeneous if the distribution of increments $\xi(t) - \xi(s)$ depends only on *t-s* when $s \le t$. If $\xi(t)$ is a homogeneous process with independent increments, then we will assume that $0 \le t < \infty$, and it is often assumed that $\xi(0) = 0$.

Let $X$ be a random variable with cumulative distribution function $F_X$. The moment generating function (*MGF*) of $X$ (or $F_X$), denoted by $M_X(t)$, is:

$$M_X(t)=\boldsymbol{E}[e^{tX}] \tag{0.1}$$

provided this expectation exists for $t$ in some open neighborhood of 0. That is, there is an $h>0$ such that for all $t$ in $-h<t<h$, $\boldsymbol{E}[e^{tX}]$ exists. If the expectation does not exist in an open neighborhood of 0, we say that the moment generating function does not exist.

If $X$ is a continuous random variable, the following relation between its moment-generating function $M_X(t)$ and the two-sided Laplace transform of its probability density function $f_X(x)$ holds: $M_X(t)=L\{f_X\}(-t)$, since the probability density function two-sided Laplace transform is given as $L\{f_X\}(s)=\int_{-\infty}^{\infty} e^{-sx} f_X(x)dx$ and the moment-generating function's definition expands (by the law of the unconscious statistician) to:

$$M_X(t)=\boldsymbol{E}[e^{tX}]=\int_{-\infty}^{\infty} e^{tx} f_X(x)dx\,. \tag{0.2}$$

The scaled cumulant generating function (*SCGF*) (see Eq. (7)) $k(r)$ of the process $\zeta(t)$ Eq. (2) is:

$$k(r)=t^{-1}\ln \boldsymbol{E}e^{r\zeta(t)}=-cr+\lambda(\boldsymbol{E}e^{-r\zeta_1}-1)\,. \tag{0.3}$$

If $\xi(t)$ is a homogeneous stochastically continuous process with independent increments in $R^m$, then its characteristic function takes the form (a Lévy-Khinchin type representation):

$$\boldsymbol{E}e^{i(z,\,\xi(t))}=\exp\{t[\,i(z,a)-\frac{1}{2}(Bz,z)+\int_{|x|\le 1}(e^{i(z,x)}-1-(z,\,x))\Pi(dx)+\int_{|x|>1}(e^{i(z,x)}-1)\Pi(dx)]\}\,, \tag{0.4}$$

where $a\in R^m$; $B$ is a symmetric non-negative operator in $R^m$; $\Pi$ is a measure in $R^m$, for which $\int \frac{(x,x)}{1+(x,x)}\Pi(dx)<\infty$ and $\Pi(\{0\})=0$.

The characteristic function of a random variable $\xi$ with a distribution function $F(x)=P\{\xi<x\}$ is defined in expression (7).

For the characteristic function (*CF*) of the process $\xi(\theta_s)$, defined as:

$$\boldsymbol{E}e^{i\alpha\xi(\theta_s)}=s\int_{-\infty}^{\infty}\boldsymbol{E}e^{i\alpha\xi(t)}e^{-st}dt\,, \tag{0.5}$$

we consider $\xi(\theta_s)$ as a randomly stopped process [68]. Randomly stopped processes include an exponentially distributed random variable $\theta_s$ that is independent of the process $\xi(t)$, $P\{\theta_s>t\}=e^{-st},\, s>0,\, t>0$. The characteristic function of a randomly stopped process is the Laplace-Carson transform of the characteristic function given by equation (7). The dependence on time $t$ is determined by applying the inverse Laplace-Carson transform $L^{-1}$. Since $\boldsymbol{E}(\theta_s)=1/s$, the parameter s can be interpreted as the average inverse random time.

Some boundary functionals considered below describe physical quantities characterizing time. These quantities are positive, and their characteristic functions transform into the Laplace transform, which have a probabilistic meaning. For example, for the Laplace transform $\boldsymbol{E}e^{-\gamma\tau^+}=\int_0^{\infty} e^{-\gamma\tau^+} f(\tau^+)d\tau^+$ of the quantity $\tau^+$, the moment of the first exit for the level $x>0$, Eqs. (5)-(6), the equality $\boldsymbol{E}e^{-\gamma\tau^+}=P\{\tau^+\le\tau\}$, $P\{\tau>t\}=e^{-\gamma t}$, is true, where $\tau$ is a random variable independent of $\xi=\tau^+$ and having an exponential distribution with parameter $\gamma$. Matching against a randomly stopped process (0.5) results in a match of the form: $s\leftrightarrow\gamma,\, \theta_s\leftrightarrow\tau$.

For the characteristic function (0.5) of $\xi(\theta_s)$ the following relation is satisfied:

$$\varphi(s,\alpha) := \boldsymbol{E} e^{i\alpha\xi(\theta_s)} = \frac{s}{s-\Psi(\alpha)},$$

and also, the fundamental factorization identity, which is also the identity of infinitely divisible factorization:

$$\varphi(s,\alpha) = \varphi_+(s,\alpha)\varphi_-(s,\alpha), \qquad \varphi_\pm(s,\alpha) = \boldsymbol{E} e^{i\alpha\xi^\pm(\theta_s)},$$

where $\zeta^\pm(t) = \sup_{0\le t'\le t}(\inf)\zeta(t')$ represents the extremes of the process $\zeta(t')$ on the interval [0, *t*]. These relationships are used in Section 5.5.

Some boundary functionals of the random process are defined in Section 2 by formulas (5)-(6), (9)-(11). Let us also define additions to the process extrema on the interval [0, *t*]:

$$\bar{\xi}(t) = \xi(t) - \xi^+(t), \quad \tilde{\xi}(t) = \xi(t) - \xi^-(t). \tag{0.6}$$

The moment of the first jump through the level $x>0$ and the values of the first undershoot and the jump that covers the level $x>0$ is defined in expression (5). The moment of the first achievement of the maximum is denoted by:

$$T^+(t) := \inf\{u \ge 0 : \xi(u) = \xi^+(t)\}. \tag{0.7}$$

This functional is discussed in Section 5.2, formula (57).

Functionals related to crossing or reaching the upper level are designated by the index "+". Let's call them upper functionals. Lower functionals related to crossing the lower-level $x<0$ is designated similarly:

$\tau^-(x) = \inf\{t : \xi(t) < x\}, \quad \gamma^-(x) = x - \xi(\tau^-(x)),$

$\gamma_-(x) = \xi(\tau^-(x) - 0) - x, \quad \gamma^-{}_x(x) = \gamma^-(x) + \gamma_-(x),$

$T^-(t) = \inf\{u \ge 0 : \xi(u) = \xi^-(t)\}.$

## 2. Definitions of some boundary functionals

In book [12], alongside the First-Passage Time (*FPT*), several other boundary functionals are discussed: the distribution of extreme values of a process, the distribution of time spent above a given level, the return time of a function within an interval, and more. The significance of these distributions for risk theory is also defined in works [7, 16, 19-21, 67].

### 2.1. On the physical meaning of risk theory

In general, risk theory applied to physical problems is associated with numerous issues explored in physical tasks involving the first-passage time of a certain level. However, the application of risk theory entails a more detailed examination and analysis of these issues.

In guides on risk theory, one can find discussions regarding life insurance models. As outlined in Reference [12], we derive from the overarching theory of stochastic processes, considering risk theory as a subset of this general framework. This article employs elements of risk theory that are related to the theory of stochastic processes and possess a common mathematical foundation.

The mathematical theory of risk, rooted in stochastic process theory, is predominantly utilized in problems related to insurance, bankruptcy assessments, and associated phenomena. Nevertheless, risk theory is a mathematical discipline closely linked to queue theory and stochastic storage theory. Due to the generality of the mathematical theory, it can also be applied in other scientific fields, such as physics, chemistry, and biology. Its research subjects may include systems at certain critical values. For instance, the shell of a nuclear reactor is designed for operation within a specific time interval even before it commences operation. The expiration of this period can be regarded as analogous to bankruptcy in the financial domain. The operational lifespan of the reactor shell may be extended based on experimental studies of the material composing the shell. Alternatively, an annealing procedure may be conducted, wherein the shell is placed in a high-temperature environment that mitigates material defects. Various processes related to the end of the operational lifespan, its extension, and the restoration of functionality may serve as subjects for investigation within risk theory. The physical problems associated with risk theory are extensive and diverse. The first-passage time of a level, one of the most utilized boundaries functionals, is employed in risk theory and has numerous physical and other applications.

### 2.2. Risk process

The actuarial risk models serve the clarification of the relation between the premiums and the surpluses during the insurance business and provide the tools for correct calculation of the risks, premiums and reserves. We assume that the insurance company begins at time 0 with initial capital $u \geq 0$ and takes a total premium up to time $t$ equal to $C(t)$. We can see that $C(t)$ represents an increasing stochastic process and in its simplest version when it increases with constant premium rate $c>0$, the function $C(t)$ obtains the form of a deterministic linear function $C(t)=ct$.

Let $N(t)=N_t$ denote the number of claims coming during the interval [0, $t$] and $\{T_k,\ k \geq 0\}$ denote the set of consecutive arrival times of the claims. Hence $N(t) = max\{k \ \geq \ 0 \ : \ T_k \ \leq \ t\}$ $= min\{k \ \geq \ 0 \ : \ T_{k+1} > \ t\}$, where by convention $T_0=0$. The size of the $k$ claim is denoted by $\xi_k$.

The primary quantity in risk theory is the classical risk process, often referred to as the reserve risk process, which takes the form:

$$R(t) = u + ct - S(t). \tag{1}$$

In expression (1) $S(t) = \sum_{k=1}^{N_t} \xi_k$, $P\{\xi_k \neq 0\} = 1$, represent the amounts of payments (claims), $\xi_k > 0$; $(\xi_k)$ is a sequence of independent identically distributed random variables with some distribution function $F(x) = P\{\xi_1 \leq x\}$, $\mu = \boldsymbol{E}\xi_1$ $(F(0)=0,\ \mu<\infty)$, $N = (N_t)_{t\geq 0}$ is the Poisson process, $N_t = \sum_k I(T_k \leq t)$, where $I(T_k \leq t) = \begin{cases} 1 & T_k \leq t \\ 0 & T_k > t \end{cases}$; $(T_{k+1} - T_k)_{k\geq 1}$ are independent variables that have an exponential distribution with parameter λ: $P\{T_{k+1} - T_k \geq t\} = e^{-\lambda t}$. The average value of $R(t)$ is $\boldsymbol{E}R(t) = u + (c - \lambda\mu)t = u + \rho\lambda\mu t$ with coefficient $\rho = c/(\lambda\mu) - 1$.

The assumptions formulated in model Eq. (1) can naturally be weakened, and the model itself can be complicated. For example, one can assume that the risk process has the form $R(t) = u + (ct + \sigma B_t) - \sum_{k=1}^{N_t} \xi_k$, where $(B_t)$ is a Brownian motion, and $(N_t)$ is a Cox process (i.e., a "counting process" with random intensity).

Sometimes, it is more convenient to use the equation:

$$\zeta(t) = u - R(t) = S(t) - ct, \quad \zeta(0) = 0, \quad c > 0, \tag{2}$$

which is referred to as the claim surplus process.

If, in the classical risk process Eq. (1), we replace the deterministic linear function *C(t)=ct* with the stochastic process $S_1(t) = \sum_{k \le \nu_1(t)} \eta_k, \quad \eta_k > 0, \quad \eta_0 > 0$, $\nu_1(t)$ is a simple Poisson process independent of $N_t$ with intensity $\lambda_1$, then:

$$R(t) = u + S_1(t) - S(t), \quad \zeta(t) = S(t) - S_1(t). \tag{3}$$

The processes *R(t)* and *ζ(t)* are called the reserve and claim surplus risk processes with random premiums, respectively. The scaled cumulant generating function (*SCGF*) of the process *ζ(t)*, Eq. (2), (see expression (7)) is given by $k(r) = t^{-1} \ln \boldsymbol{E} e^{r\zeta(t)} = -cr + \lambda(\boldsymbol{E} e^{-r\xi_1} - 1)$, Eq. (0.3).

Furthermore,

$$R(t) = u + S_1(t) - S(t), \quad u > 0, \quad S_1(t) = \sum_{k \le \nu_1(t)} \eta_k, \quad S(t) = \sum_{k=1}^{N_t} \xi_k, \quad P\{\xi_k \ne 0\} = P\{\eta_k \ne 0\} = 1. \tag{4}$$

If the premiums $\eta_k$ have the characteristic function $\varphi_1(\alpha) = c(c - i\alpha)^{-1}$, then $R(t) = S_1(t) - S(t)$ is an upper almost continuous risk process with exponentially distributed premiums and initial capital *u*=0. Such processes with random premiums include the fluctuations of trajectory observables, as discussed in work [62]. These issues are considered in greater detail in Refs. such as [7, 12, 19, 21].

Let us dwell briefly on the question of the nature of the distributions *F=F(x)* of the amounts of payments. Events associated with payments are usually (albeit rather conditionally) attributed to one of three types: a) normal, b) extreme, c) catastrophic. To describe normal events, distributions with rapidly decreasing "tails" are used (for example, an exponential distribution with the condition $1-F(x) \sim e^{-x}, x \to \infty$). To describe "extreme" events, distributions *F=F(x)* with "heavy tails" are used, for example, $1-F(x) \sim x^{-\alpha}, x \to \infty, \alpha > 0$ (Pareto-type distribution) or $1 - F(x) = \exp\{-(\frac{x-\mu}{\sigma})^p\}, x > \mu$ (Weibull distribution) with $p \in (0,1)$.

In the context of risk theory, instead of representing initial capital, the value *u* in equations (1)-(4) can also hold a physical interpretation, such as representing energy (for example, the internal energy of a statistical system).

### 2.3. Boundary functionals

Let us indicate what boundary functionals are. In book [11], it is written: "Most problems for such processes are related to the so-called boundary problems, in which boundary functionals of process trajectories are studied, associated with the exit of the process trajectory beyond a certain level. In this case, two types of boundaries are distinguished: absorbing and repulsive. The intersection of the trajectories of the absorbing boundary process determines the boundary functionals from the trajectory of the process to the moment of intersection. In particular, the moment of intersection of a given level also serves as a boundary functional. Repulsive boundaries "distort" the trajectory of the process, i.e., change its initial characteristics, and, consequently, the boundary functionals determined by another absorbing level." In book [13], it is written: "... boundary functionals (i.e., functionals from selective trajectories associated with the exit of these trajectories beyond certain boundaries)".

Below, various random times will be defined. Following the formal definition (book [10]), we will also provide an informal definition of random time. Let us consider stochastic experiments conducted continuously over time, and let event *A* be such that its occurrence can be determined by observing the results of the experiments up to some random moment in time. Such a moment in time will be referred to as a random time. It is also sometimes referred to as a random variable that is independent of the future, a Markov moment, or a stopping time.

For certain functionals of the random process *ξ(t),* definitions are provided in books [12, 67]. These include:

$\tau^+(x) = \inf\{t : \xi(t) > x\},\ \ x > 0$ is the moment of the first exit for the level *x*>0; (5)

$\tau^-(x) = \inf\{t : \xi(t) < x\},\ \ x < 0$ is the moment of the first exit for the level *x*<0;

$\gamma^+(x) = \xi(\tau^+(x)) - x$ is first overjump over *x*>0;

$\gamma_+(x) = x - \xi(\tau^+(x) - 0)$ is the first under jump over *x*>0;

${\gamma^+}_x(x) = \gamma^+(x) + \gamma_+(x)$ is the first jump that covers *x*>0;

$\gamma^-(x) = \xi(\tau^-(x)) - x$ is first overjump over *x*<0.

In risk theory, $\tau^+(u) = \tau(u)$ is defined as the ruin time. In physics, this is the moment when the value *R(t)* from equation (1) reaches zero, representing the moment of degeneration or the end of the lifetime, also referred to as the first-passage time (*FPT*). This is related to the fact that in physical problems, the attainment of a zero level by a certain process is often referred to as the end of the lifetime of the quantity described by that process, or as the *FPT* of the zero level. In risk theory, this is not the case; the system can recover, as in queueing theory, where idle periods of downtime and busy alternate with one another. Physical analogues of such phenomena also exist. For the claim surplus process Eq. (2), this is the *FPT* when the process *ζ(t)* reaches the level *u*:

$$\tau(u) = \inf\{t : R(t) < 0\}, \quad \tau^+(u) = \inf\{t : \zeta(t) > u\}. \tag{6}$$

The characteristic function of a random variable $\xi$ with a distribution function $F(x) = P\{\xi < x\}$ is referred to as a complex-valued function:

$$\varphi(t) = \boldsymbol{E}e^{it\xi(t)} = \int_{-\infty}^{\infty} e^{itx} dF(x).$$

A process with independent increments $\xi(t)$ is called homogeneous if it is defined on $[0,\infty)$, $\xi(0) = 0$, and the distribution $\xi(t+h) - \xi(t)$ coincides with the distribution $\xi(h)$ for all $t > 0$ and $h > 0$. The characteristic function of a homogeneous process $\xi(t),\ t \geq 0$ is determined in the theory of random processes (for $\xi(0) = 0$) (works [10, 12, 17, 18]) by the relation:

$$\boldsymbol{E}e^{i\alpha\xi(t)} := \int_{-\infty}^{\infty} e^{i\alpha x} dF(x) = e^{t\Psi(\alpha)},\ \ t \geq 0, \tag{7}$$

where $F(x) = P(\xi < x)$ is the distribution function of a random process $\xi(t),\ t \geq 0$, and the function $\psi(\alpha)$ represents the scaled cumulant generating function (*SCGF*) of the process $\xi(t),\ t \geq 0$. The *SCGF* do not depend on *t* and characterize all finite-dimensional distributions of the process.

If for a process $\xi(t),\ t \geq 0$, the function $\psi(\alpha)$ at $i\alpha = r$ is equal to *s*, then we obtain the equation:

$$\Psi(\alpha)\big|_{i\alpha=r} := k(r) = s, \quad \pm \operatorname{Re} r \geq 0. \tag{8}$$

This equation, in risk theory (works [12, 19, 21, 67]), is referred to as the fundamental Lundberg equation.

We will now rewrite the definitions from Eq. (5) using the notations from work [67]:

$$Y^+(u) = -R(\tau(u)), \quad \gamma^+(u) = \zeta(\tau^+(u)) - u, \tag{9}$$

$$X^+(u) = R(\tau(u) - 0), \quad \gamma_+(u) = u - \zeta(\tau^+(u) - 0),$$

$$X^+(u)+Y^+(u)=R(\tau(u)-0)-R(\tau(u)),$$
$$\gamma^+{}_u=\gamma^+(u)+\gamma_+(u).$$

Here, $Y^+(u)=\gamma^+(u)$ denotes the severity of ruin, $X^+(u)=\gamma_+(u)$ is the value of the process $R(t)$ just before the onset of bankruptcy, referred to as the surplus prior to ruin, $\gamma^+{}_u$ is the amount of the claim causing ruin; *x*=*u*, *x* from Eq. (5), *u* from Eq. (9).

Additionally:

- $\zeta^\pm(t)=\sup_{0\le t'\le t}(\inf)\zeta(t')$ represents the extremes of the process $\zeta(t')$ on the interval [0, *t*].
- $\zeta^\pm=\sup_{0\le t<\infty}(\inf)\zeta(t)$ represents the absolute extremes of the process $\zeta(t)$.
- $\tau'(u)=\inf\{t>\tau(u),\ R(t)>0\}$ is the moment of return $R(t)$ after bankruptcy to the half-plane $\Pi^+=\{y>0\}$.

Let us also denote:

$$T'(u)=\begin{cases}\tau'(u)-\tau(u), & \tau(u)<\infty\\ \infty, & \tau(u)=\infty\end{cases}. \tag{10}$$

This is called the "red time," which determines the duration of stay of the process $R(t)$ in the half-plane where $\Pi^-=\{x<0\}$, the amount of time the risk reserve process $R(t)$ stays below zero after the ruin time $\tau(u)$, work [67], the interval between moments $\tau'(u)$ and $\tau(u)$, and the process *ζ(t)* is below the level *u*;

$$Z^+(u)=\sup_{\tau^+(u)\le t<\infty}\zeta(t)=\sup_{\tau(u)\le t<\infty}\{-R(t)\}, \tag{11}$$
$$Z_1^+(u)=\sup_{\tau^+(u)\le t<\tau'(u)}\zeta(t)=\sup_{\tau^+(u)\le t<\tau'(u)}\{-R(t)\},$$

$Z^+(u)$ defines the total maximal deficit, $Z_1^+(u)$ the maximal deficit on $T'(u)$; $Z^+(u)$ defines the total maximal deficit, which is the largest shortfall encountered by the process $T'(u)$ during its lifetime. In other words, $Z^+(u)$ is the most negative value (or largest deficit) reached by the process when it goes below zero. Similarly, $Z_1^+(u)$ defines the maximal deficit on a finite time interval [0, $\tau$`($u$)]. This is the largest deficit encountered by the process $T'(u)$ during the time interval [0, $\tau$`($u$)]. There are other boundary functionals, for example Eqs. (0.6)-(0.7).

All of these quantities, in addition to their values in risk theory, also have physical interpretations. For example, in problems involving achieving a certain particle threshold necessary for a reaction or biochemical process, the value of the jump over a given level represents the excess beyond this threshold. In such cases, the conditions for the reaction or biochemical process may change. The moment of bankruptcy corresponds to the first passage time (*FPT*). In chemical reactions, the time spent above the threshold determines whether there is sufficient time for the reaction to occur, provided that the number of molecules (level *x*=*u*) is adequate. An insurance company can continue operating after bankruptcy by taking out a loan, and distributions of jump functionals are used to determine the loan amount. There are also physical analogs for these quantities.

It is also possible to indicate the physical meaning of the probabilistic interpretation of the Laplace transform, noted between expressions (0.5) and (0.6). The Laplace transform $\boldsymbol{E}e^{-\gamma\xi}$ at $\gamma\ge 0$ represents the probability that the moment of reaching a given level *x* by the process $\xi$ will occur before the moment of stopping the observation *τ*, which has an exponential distribution.

## 3. Characteristic functions (*CF*) and *MGF* of boundary functionals

In works [7, 12, 67], expressions for the distributions of the functionals specified in Section 2 were obtained.

Following work [12], we consider the case of the risk process being in the survival and risk zones. We examine the claim surplus risk process Eq. (2) with an initial capital $u$>0. We assume $m = \boldsymbol{E}\zeta(1) = \frac{\lambda - cb}{b} < 0$; parameters $\lambda$, $c$ is defined in Eqs. (1)-(2), parameter $b$ in Eq (12). The moment $\tau^+(u) = \tau(u)$ Eq. (6) determines not only the time of bankruptcy (i.e., the first exit $\zeta(t)$ into the "red" zone, $\{y>u\}$, the risk zone), but also the duration of the initial stay in the survival zone $\{y \le u\}$ , which can be conditionally referred to as the "green" zone. The value $\tau'(u) = \inf\{t > \tau^+(u), \zeta(t) < u\}$ determines the moment of the first return to this zone. The duration of the first stay $\zeta(t)$ in the "red" zone is determined by the value Eq. (10). The probability of bankruptcy, which in risk theory is denoted by $\Psi(u)$, is given by the "tail" of the distribution $\zeta^+$ , $\Psi(u) = P\{\zeta^+ > u\} = \bar{P}_+(u)$, book [12].

### 3.1. Approximation used. Lundberg equation

We focus on the example of the claim surplus risk process Eq. (2) with an exponential distribution of claims $\xi_k$, $S(t) = \sum_{k \le \nu(t)} \xi_k$, $P\{\xi_k > 0\} = 1$, as in the Kramers-Lundberg model (Refs. [7, 12]), where:

$$\bar{F}(x) = P\{\xi_k > x\} = e^{-bx}, \quad x > 0, \quad \mu = \boldsymbol{E}\xi_k = b^{-1}, \quad b > 0. \tag{12}$$

It is possible to consider the general case, but it leads to cumbersome expressions that make it difficult to derive an explicit final result. Therefore, we restrict our analysis to this example. In Section 5.6, some other distributions of the magnitude of process jumps are considered. In Section 5.3, a comparison is made between the results obtained using approximation (12) and without it.

The moment generating function (*MGF*, Eqs. (0.1), (0.2)) of the process $\zeta(t)$ (or $\zeta(\theta_s)$, see the definition of a randomly stopped process in Eq. (0.5)), is determined by the *SCGF* Eq. (7) at $\Psi(\alpha)\big|_{i\alpha=r} = k(r)$ , as in Eq. (8), $k(r) = t^{-1} \ln \boldsymbol{E}e^{r\zeta(t)} = -cr + \lambda(\boldsymbol{E}e^{-r\zeta_1} - 1)$ , Eq. (0.3), $\boldsymbol{E}e^{-r\zeta_1} = b/(b-r)$, $b$>$r$, Eq. (12) where:

$$\boldsymbol{E}e^{r\zeta(t)} = e^{tk(r)}, \quad \boldsymbol{E}e^{r\zeta(\theta_s)} = \frac{s}{s - k(r)}, \tag{13}$$

$$k(r) = \frac{\lambda r - cr(b-r)}{b-r}, \quad m = \boldsymbol{E}\zeta(1) = \frac{\partial k(r)}{\partial r}_{|r=0} = \frac{\lambda - cb}{b} < 0.$$

The Lundberg equation (8) can be derived by setting the denominator in equation (13) equal to zero. In the case of the function $k(r)$ given by Eq. (0.3) and Eq. (13), the Lundberg equation Eq. (8) simplifies to a quadratic equation (for $r$<$b$):

$$s - k(r) = 0 \sim cr^2 + (s + mb)r - sb = 0. \tag{14}$$

This equation has two roots; $r_1 = -\rho_-(s)$, $r_2(s) = \rho_+(s) > 0$ :

$$\rho_+(s) = \frac{1}{2c}[-(s+mb) + \sqrt{(s+mb)^2 + 4scb}], \quad \rho_-(s) = \frac{1}{2c}[(s+mb) + \sqrt{(s+mb)^2 + 4scb}].$$

Knowledge of roots $\rho_-(s),\ \rho_+(s)$ is important for the study of boundary functionals.

### 3.2. Distribution of extreme values of the process. Extremes of the claim surplus process

The roots $r_1 = -\rho_-(s)$ and $r_2(s) = \rho_+(s) > 0$ of equation (14) determine the *MGF* of the extrema $\zeta^{\pm}(\theta_s)$ of the process Eq. (2) and their distributions:

$$\boldsymbol{E}e^{-z\zeta^+(\theta_s)} = \frac{p_+(s)(b+z)}{\rho_+(s)+z}, \quad \rho_+(s) = bp_+(s), \tag{15}$$

$$\boldsymbol{E}e^{-z\zeta^-(\theta_s)} = \frac{\rho_-(s)[1-\rho_-(s)/b]}{\rho_-(s)+z}, \quad cp_+(s)\rho_-(s) = s.$$

Here, $b$ are the parameters (indices) of the exponential distributions (in accordance with Eq. (12)) of claims (jumps $\xi_k$ in the process $S(t)$ from equation (2)), and $c$ represents the intensity of insurance premium receipts, as described in Eqs. (1)-(2). The probability of reaching the minimum extremum is given by work [12]:

$$P_-(s,x) = P\{\zeta^-(\theta_s) < x\} = e^{\rho_-(s)x}, \quad x \le 0.$$

The probability of the maximum extremum is given by:

$$\bar{P}_+(s,x) = 1 - P_+(s,x),\ \ P_+(s,x) = P\{\zeta^+(\theta_s) < x\} = 1 - E\{e^{-s\tau^+(x)}, \tau^+(x) < \infty\},$$

$$\bar{P}_+(s,x) = q_+(s)e^{-\rho_+(s)x}, \quad x > 0, \quad p_+(s) = P\{\zeta^+(\theta_s) = 0\} > 0, \quad q_+(s) = 1 - p_+(s).$$

If $m = dk(r)/dr\big|_{r=0} < 0$, then $\rho_-(s)_{s\to 0} \to 0,\ \rho'_-(0) = \dfrac{1}{|m|},\ \rho_+(s)_{s\to 0} \to \dfrac{b|m|}{c}$,

the probability of bankruptcy, which in risk theory is denoted by $\Psi(u)$, is given by work [12]:

$$\bar{P}_+(s,u)_{s\to 0} \to \Psi(u) = q_+e^{-\rho_+u}, \quad u>0, \quad q_+(s=0) = q_+, \quad \rho_+(s=0) = \rho_+. \tag{16}$$

Expression (16) describes the Kramers-Lundberg approximation (in the approximation [12]) for the probability of bankruptcy (surplus-level process).

### 3.3. Characteristic function (*CF*) and *MGF* of other boundary functionals

Taking into account that for model Eq. (12) $q_+ = \dfrac{\lambda}{cb},\ \rho_+ = bp_+ = \dfrac{bc-\lambda}{c}$, the *MGF* of $\tau^+(u),\ \tau'(u),\ \gamma^+(u),\ T'(u)$, Eqs, (5)-(6), (9)-(10), are determined by books [12, 67]:

$$q_+(s,u) = \boldsymbol{E}[e^{-s\tau^+(u)}, \tau^+(u) < \infty] = q_+(s)e^{-\rho_+(s)u}, \quad u \ge 0, \tag{17}$$

$$u = 0,\ m < 0,\ q_+(s) = \boldsymbol{E}[e^{-s\tau^+(0)}, \tau^+(0) < \infty] = \frac{\lambda}{c}\int_0^{\infty} e^{-z\rho_-(s)}\bar{F}(z)dz = \frac{\lambda}{c}\frac{1}{\rho_-(s)+b},$$

$$\lim_{s\to 0} q_+(s,u) = \Psi(u) = q_+e^{-\rho_+u},\ u \ge 0,$$

$$\varphi_u(s) = \boldsymbol{E}[e^{-s\tau'(u)}, \tau^+(u) < \infty] = q_+(s,u)\frac{b}{b+\rho_-(s)},$$

$$\tilde{q}_+(z,u) = \boldsymbol{E}[e^{-z\gamma_+(u)}, \tau^+(u) < \infty] = \Psi(u)\frac{b}{b+z},$$

$$\tilde{q}_u(s) = \boldsymbol{E}[e^{-sT'(u)}, \tau^+(u) < \infty] = \Psi(u)\frac{b}{b+\rho_-(s)}.$$

From the last two formulas, it follows that $\gamma_+(u)$ and $T'(u)$ do not depend on *u*. This followed from Eqs. (16)-(17) when $\boldsymbol{E}[\tau^+(u) < \infty] = q_+ e^{-\rho_+ u} = \Psi(u)$. Therefore, only for this example of the exponential distribution Eq. (12), we can express the *MGF* as:

$$\boldsymbol{E}[e^{-z\gamma_+(u)} / \tau^+(u) < \infty] = \boldsymbol{E}e^{-z\tilde{\gamma}_+(u)} = \frac{b}{b+z}, \tag{18}$$

$$\boldsymbol{E}e^{-s\tilde{T}'(u)} = \boldsymbol{E}e^{-\rho_-(s)\tilde{\gamma}^+(u)} = \frac{b}{b+\rho_-(s)}, \quad u > 0;$$

$$\tilde{q}_+(z,0) = \varphi(0)\frac{b}{b+z}, \quad \tilde{q}(s) = \Psi(0)\frac{b}{b+\rho_-(s)}, \quad for \quad u = 0,$$

$$\varphi(s) = q_+(s)\frac{b}{b+\rho_-(s)}, \quad \varphi(0) = \tilde{q}(0) = \Psi(0) = q_+ .$$

Let us denote the following:

- $\theta_{N_u} = \theta^{(u)}{}_{N_u}$ is the total duration of survival periods (by analogy with queueing theory), in the survival zone $\{y \le u\}$;

$$\sigma_{N_u} = \sigma^{(u)}{}_{N_u} \tag{19}$$

- is the total duration of "red" periods, in the risk zone {*y*>*u*} (as defined in Eq. (10));
- $S_{N_u} = S^{(u)}{}_{N_u}$ is the total duration of regeneration periods, when $\zeta(t) < u$.

The moment of the first regeneration of the claim surplus process $\zeta(t)$ is given by:

$$\tau'_1(u) = \inf\{t : t > \tau^+(u) : \zeta(t) < u\} = \tau^+{}_1(u) + T'_1(u), \tag{20}$$

$$\theta^{(u)}{}_n = \tau^+{}_1(u) + \sum\nolimits_{k=2}^{n} \tau^+{}_k(0), \ (\theta_n = \sum\nolimits_{k=1}^{n} \tau'_k(0), \ u = 0),$$

$$\sigma^{(u)}{}_n = T'_1(u) + \sum\nolimits_{k=2}^{n} T'_k(0), \ (\sigma_n = \sum\nolimits_{k=2}^{n} T'_k(0), \ u = 0),$$

where $T`_k$ and $\tau^+{}_k$ describe *k*-th regeneration. The value $S_{N_u}$ is determined through a random walk $S_n = \sum\nolimits_{k=0}^{n} \xi_k, \quad S_0 = 0$, described by the sequence:

$$\{\tau'_k(0)\}_{k\ge1} : S_n = \sum\nolimits_{k\le n} \tau'_k(0), u = 0,$$

$$\{\tau'_k(u)\}_{k\ge1} : S^{(u)}{}_n = \tau'_1(u) + \sum\nolimits_{k=2}^{n} \tau'_k(0), u > 0.$$

The renewal process determines the number of renewals on the interval [0, *t*] for $S_n$, which we denote by:

$$N(t) = \max\{n : S_n < t\}, \ H(t) = \boldsymbol{E}N(t),$$

$N(\theta_s)$ is a randomly stopped process $N(t)$, Eq. (0.5).

In book [12], the distributions for $N_u(\theta_s)$ and $N_u(N)$ are found, and subsequently, from the distribution of $N_u$, the *MGF*s can be expressed:

$$\boldsymbol{E}[e^{-s\sigma_{N_u}}, \sigma_{N_u} < \infty] = 1 - \Psi(u) + \frac{p_+\Psi(u)}{1-q_+q_+(s)}\frac{b\Psi(u)}{b+\rho_-(s)}, \tag{21}$$

$$\boldsymbol{E}[e^{-s\theta_{N_u}}, \theta_{N_u} < \infty] = 1 - \Psi(u) + \frac{p_+\Psi(u)}{1-q_+q_+(s)} q_+(s) e^{-\rho_+(s)u},$$

$$\boldsymbol{E}[e^{-sS_{N_u}}, S_{N_u} < \infty] = 1 - \Psi(u) + \frac{p_+\Psi(u)}{1 - q_+ q_+(s)} \frac{bq_+(s)}{b + \rho_-(s)} e^{-\rho_+(s)u},$$

$$\boldsymbol{E}e^{-s\tilde{\sigma}_{N_u}} = 1 - \frac{\rho_-(s)\Psi(u)}{\rho_-(s) + \rho_+}\Big|_{s\to 0} \to 1, \quad \frac{q_+\rho_-(s)}{\rho_-(s)+\rho_+} = \frac{\rho_+(s) - \rho_+}{\rho_+(s)},$$

where unlike Eq. (19) $\tilde{\sigma}_n = \sum_{k \le n} [T'_k(u) / \tau^+(u) < \infty]$.

From Eq. (21) at $s \to 0$ it follows that

$$P\{\sigma_{N_u} < \infty\} = \ldots = 1 - \Psi(u) + \frac{\Psi^2(u)}{1 + q_+},$$

$$P\{\sigma_N < \infty\} = \frac{1}{1 + q_+} = \frac{cb}{\lambda + cb} \quad (\lambda \le cb).$$

The density of the marginal bankruptcy functions (for *u*=0) is determined by the following relations:

$$p_1(s,0,x) = \frac{\lambda b}{c} \frac{e^{-bx}}{b + \rho_-(s)} = bq_+(s)e^{-bx}, \tag{22}$$

$$p_2(s,0,x) = \frac{\lambda}{c} e^{-(b+\rho_-(s))x} \ne p_1(s,0,x),$$

$$p_3(s,0,x) = \frac{\lambda b}{c\rho_-(s)} (1 - e^{-\rho_-(s)x}) e^{-bx}, \quad q_+(s) = \frac{\lambda}{c} \frac{1}{b + \rho_-(s)},$$

where in general case, for any distributions *F(x)*:

$$p_1(s,0,x) = \frac{\lambda}{c} \int_0^\infty e^{-\rho_-(s)y} dF(x+y),$$

$$p_2(s,0,x) = \frac{\lambda}{c} e^{-\rho_-(s)x} \bar{F}(x) \ne p_1(s,0,x),$$

$$p_3(s,0,x) = \frac{\lambda}{c\rho_-(s)} (1 - e^{-\rho_-(s)x}) F'(x), \quad x > 0,$$

are the densities of distributions of quantities $\gamma^+(0), \gamma_+(0)$ from expressions (9), in Eq. (22) $\bar{F}(x) = P\{\xi_k > x\}$, $F(x) = P\{\xi_k < x\} = 1 - e^{-bx}$. From the pre-boundary marginal bankruptcy functions $\bar{P}_k(s,u,x) = P\{\zeta^+(\theta_s) > u, \gamma_k(u) > x\}$, ( $\gamma^+(u) = \gamma_1(u), \gamma_+(u) = \gamma_2(u), \gamma_u{}^+(u) = \gamma_3(u)$ ), (integrals over *x* from Eq. (22) and more general expressions written below Eq. (22)) we obtain for *x*=0 the relation $q_+(s) = \frac{\lambda}{c} \frac{1}{b + \rho_-(s)}$.

The *MGF* of the quantity $Q_x(\infty)$, where $Q_x(t) = \int_0^t I\{\xi(u) > x\} du$, $I\{A\} = \begin{cases} 1, & x \in A \\ 0, & x \notin A \end{cases}$, representing the time spent above the level *x* in a semi-continuous as well as an almost semi-continuous from above process $\zeta(t)$, is given by the relation:

$$D_x(\mu) = \boldsymbol{E}e^{-\mu Q_x(\infty)} = 1 - \frac{\rho_+(\mu) - \rho_+}{\rho_+(\mu)} e^{-\rho_+ x}, \; x > 0. \tag{23}$$

This agrees with the last relation from Eq. (21) for $x=u$, i.e., $Q_u(\infty)=\tilde{\sigma}_{N_u}$ for Eq. (12). Furthermore, the integral transformation $Q_x(\infty)$ of the time spent in the risk zone $\{y>u\}$ is also defined.

## 4. Example. Unicyclic network with affinity A, an asymmetric random walk

### 4.1. Statement of the problem. Approximations used.

We apply the relations developed in Sections 2 and 3 to the unicyclic network with affinity *A* (works [69-71]) in periodically controlled systems. These results are applicable, for example, to the description of the behavior of a colloidal particle controlled by a periodic field, molecular pumps, enzymatic reactions with stochastic substrate concentrations, and networks of enzymatic reactions or small electronic circuits.

Current fluctuations that are universally observed in systems can be represented in terms of continuous-time Markov networks with a scaled cumulant generating function $\lambda(z)$, Eqs. (24) and (25).

A unicyclic graph is a type of graph that contains exactly one cycle, which is a closed path of edges that begins and ends at the same vertex. In other words, a unicyclic graph is a connected graph with exactly one cycle. It can have any number of vertices and edges, as long as it satisfies this condition.

"Unicyclic networks" are graphs obtained from a binary leaf labeled as a ("phylogenetic") tree by adding an edge to introduce a cycle. This class of graphs provides a useful representation of hybrid evolution in molecular evolutionary biology.

In Ref. [69], a Markovian network consisting of $N$ discrete states $\{i\}$ is considered, allowing for transitions with rates $k_{ij}\geqslant 0$ from state $i$ to state $j$. All transitions are assumed to be reversible, meaning that $k_{ij}>0$ implies $k_{ji}>0$. In work [73], a complete set of fundamental cycles $\{\beta\}$ within the network is identified. Each cycle is associated with an affinity $A_\beta$, a fluctuating current $X_\beta(t)$ that counts the number of cycle completions after time $t$ (the so-called integrated current), and an average current $J_\beta \equiv \langle X_\beta(t)\rangle / t$, where the brackets indicate an average over stochastic trajectories. This average is independent of $t$ for initial conditions drawn from the steady-state distribution.

A generalized distance $d^\beta_{ij}=-d^\beta_{ji}$ is introduced, which quantifies how much $X_\beta$ increases during the transition from $i$ to $j$. These increments are constrained to sum to one for each closed loop that completes the cycle once in the forward direction. For example, $X_\beta$ could represent the scaled displacement of a molecular motor. In this context, the affinity $A_\beta$ would be given by the external force multiplied by the length of the full step of the motor, while $d^\beta_{ij}$ denotes the relative length of the substep associated with the conformational change from $i$ to $j$.

Adopting a vector notation $\vec{X}$ for the set of all cycle currents $X_\beta$ the scaled cumulant generating function is defined as:

$$\lambda(\vec{z}) \equiv lim_{\,t\to\infty}\ (1/\ t)\ ln\, \mathbf{E}exp\left[\vec{z}\cdot\vec{X}(t)\right], \tag{24}$$

where $\vec{z}$ is a real vector.

Comparing Eq. (24) with equations (7), (8), and Eq. (13), we see that $\lambda(\vec{z})$ coincides with the function $k(\vec{r})$ when $\vec{z}=\vec{r}$, which appears in equations (8), (13)-(14).

For unicyclic networks, where there is only a single affinity $A \equiv A_\alpha$ and a single fluctuating current $X \equiv X_\alpha$, we consider a network consisting of a single cycle with *N* vertices and affinity *A*, as illustrated in Fig. 1c (Fig. 1a in Ref. [69]).

The transition rates for an arbitrary unicyclic model with *N* states and periodic boundary conditions are denoted by $k_{i,i+1} = k_i^+$ and $k_{i,i-1} = k_i^-$, where *i*=1,2, . . . ,*N*. A fixed affinity *A* imposes the constraint $(\prod_{i=1}^N k_i^+ / \prod_{i=1}^N k_i^-) = e^A$ on the transition rates. Different choices of transition rates that satisfy this condition can lead to different *SCGF*. In particular, if the transition rates are uniform, i.e., $k_i^+ = k^+$ and $k_i^- = k^-$, the *SCGF* divided by the average current $\lambda(z)/J$ becomes $\lambda_{ARW}(z, A, N)$, as given in Eq. (25).

For unicyclic networks we no longer have to distinguish between the different types of *SCGF*s and can simply write $\lambda(z) \equiv \lambda_\alpha(z)$. It has been shown in work [69] that for the unicyclic case, the *SCGF* $\lambda_\alpha(z)$ exhibits effects of faster-than-quadratic growth at large *z* and a pronounced plateau around the minimum of $\lambda_\alpha(z)$. This behavior of the *SCGF* is best illustrated using an asymmetric random walk (*ARW*). The *SCGF* is given by [72]:

$$\lambda(z) = k^+\left[e^{z/N} + e^{-(z+A)/N} - 1 - e^{-A/N}\right] = J\lambda_{ARW}(z, A, N), \quad (25)$$

$$\lambda_{ARW}(z, A, N) \equiv \{cosh\left[(z + A/2)/N\right] - cosh\left[A/(2N)\right]\}/(1/N)sinh\left[A/(2N)\right].$$

The hopping rates in the forward and backward directions are denoted by $k^+$ and $k^-$, respectively. The *SCGF* Eq. (25) is bounded by the parabola: $\lambda_{ARW}(z, A, N) \geqslant zJ(1 + z/A)$.

From expressions (15) to (23), it is evident that the characteristic function (*CF*) or *MGF* of functionals such as the time spent above a given level and the extremes of a random function are determined by the positive $\rho_+$ (and negative) roots of equations (8), (14).

In work [69], the series expansion of the exponential in $\lambda(z)$ is used. In expression (25), we expand the exponential term $exp\{-z/N\} \approx 1 - z/N$ in a series. Let's consider the case $z/N < 1$. This condition is satisfied because expression (13) is obtained under the condition $r < N$ necessary for the convergence of the integral, and below an exponential distribution of jumps with *z*=*r*, *b*=*N* is used. We represent the term $exp\{z/N\} - 1$ as:

$$\int_0^\infty (exp\{zX_\alpha\} - 1)p(X_\alpha)dX_\alpha, \; p(X_\alpha) = \delta(X_\alpha - 1/N),$$

where $X_\alpha$ is a fluctuating current that counts cycle completions after time *t* (the so-called integrated current), Ref. [69]. We replace the delta distribution density $p(X_\alpha) = \delta(X_\alpha - 1/N)$ for $X_\alpha$ with the density of the exponential distribution $p(X_\alpha) = c_5 exp\{-c_5 X_\alpha\}, c_5 = N$ and obtain expression (13) with *b*=*N*, i.e., a risk process with an exponential distribution of jumps in incoming demands. This type of process is well studied in both risk theory and stochastic storage theory (a process with an exponential distribution of input jumps). In this case, at $z = r$ expression (25) transitions to relation Eq. (13), for which functionals Eqs. (15)-(23) are derived.

Let us justify the replacement of the distribution. The clockwise jumps in the flow are equal to 1/*N*, and the distribution $p(X_\alpha) = \delta(X_\alpha - 1/N)$ is valid. However, flows fluctuate when the affinity *A* changes, and the movement can become counterclockwise, with the jump being equal to -1/*N*. This situation is described in work [73]. The exponential distribution has a maximum at zero, which corresponds to a large *N* and the proximity of the jump from 1/*N* to −1/*N*.

In Ref. [69], it is noted that the *SCGF* $\lambda(z)$ is bounded below by an exponential function of the form: $\lambda(z) \geq R\left(e^{(|\sigma/2+z\cdot J|-\sigma/2)/R}-1\right)$, where $\sigma$ is the average entropy production, $R \equiv \Sigma_i p^s{}_i r_i$ is the average number of transitions per unit time in the entire network, $p^s{}_i$ is the stationary probability distribution of state *i*, and the exit rate from state *i* is defined as $r_i \equiv \sum_l k_{il}$ .

The correspondence between the statistical properties of trajectory observables, referred to in work [69] as the integrated current of finite spatial Markov jump processes, is established by mapping *SCGF* Eq. (25) associated with these generalized currents into a Lévy-Khinchin type representation, Eq (0.4). This representation, with the special case in expression (13), corresponds to the scaled cumulant generating function of a general Lévy process.

The exponential relationship between the flow jumps $X_\alpha$ and the distribution of jumps is physically motivated by increasing the level of stochasticity, replacing the delta distribution for the flows with an exponential distribution. One might expect the reverse type of relation to hold; that is, one could inquire about generalized currents whose determining coefficients are random functions of the stationary velocities, which are associated with an exponential distribution of jumps. This would correspond to a subclass of trajectories that might be of physical interest or even have realistic applications.

This result aligns with a broad class of processes in risk theory, queue theory, and storage theory (stochastic storage systems, as previously mentioned). The results obtained using expression Eq. (25) and approximation Eq. (26) are compared in Section 5.3. It is demonstrated that the form of the dependence of the mean value on the parameter *s* remains unchanged; however, with approximation Eq. (26), the dependence is shifted.

Thus, starting from *SCGF* of the form Eq. (13):

$$k(r) = k^+[\frac{r}{b-r} - c_1 r] = \frac{\lambda r}{b-r} - cr, \quad \lambda = k^+, \quad b = N, \quad c = e^{-A/N} k^+ / N \,, \qquad (26)$$

$$m = k'(r)\big|_{r=0} = \frac{k^+}{N}[1-e^{-A/N}] < 0 \text{ at } A<0.$$

In Ref. [69], a network consisting of a single cycle with *N* vertices and affinity *A* is considered. The *SCGF* of type Eq. (26) corresponds to an example of an asymmetric random walk (*ARW*). The hopping rates in the forward and backward directions, $k^+$ and $k^-$, are uniform, with $\ln(k^+/k^-)=A/N$. The average current in this model is given by $J=(k^+-k^-)/N$, and the entropy production is $\sigma=JA$, where *J* represents the average flow.

The roots of the Lundberg equation (14), using the *SCGF* from Eq. (26), are given by:

$$\rho_+(s) = [-(s+\lambda-cb)+\sqrt{(s+\lambda-cb)^2+4scb}\,]/2c, \quad \rho_-(s) = [(s+\lambda-cb)+\sqrt{(s+\lambda-cb)^2+4scb}\,]/2c\,, \quad s>0\,. \quad (27)$$

If we take the parameter values to be $b=N=5$ (as in Fig. 1a in work [69], and Fig. 1c below), $A=-1$, $k^-=1$, then $\lambda=0.819$, $J=0.36$, $c=0.2$, and $m=-0.036$. Substituting these values into equation (27), we obtain:

$$\rho_+(s) = [-(s-0.181)+\sqrt{s^2+3.638s+0.032761}\,]/0.4, \quad \rho_-(s) = [(s-0.181)+\sqrt{s^2+3.638s+0.032761}\,]/0.4\,. \quad (28)$$

The root is positive when the "+" sign precedes the square root, and negative when the "−" sign does. These roots, $\rho_+$ (for the "+" sign) and $r_-=-\rho_-$ (for the "−" sign), satisfy the condition $m<0: \rho_-(s)_{s\to 0} \to 0,\ \rho_+(s)_{s\to 0} \to \rho_+$. The behavior of these roots, as derived from expressions Eqs. (27) and (28), is illustrated in Fig. 1a and 1b for $s>0$.

The behavior of $\rho_-(s)$ the root aligns with the phase transition at *s*=0, as observed in work [74]. In works [62, 80], it is shown that the positive root $\rho_+(s)$ in the thermodynamics of trajectories (Refs. [74–79]) corresponds to the *SCGF* $-g(s)$ of the *MGF* $Z_K(x) \equiv \int_0^\infty d\tau e^{-x\tau} P_K(\tau)$, where *τ* is the random time taken to reach a fixed value *K* of dynamic activity (in the *x*-ensemble); $P_K(\tau)$ is the distribution of total trajectory length *τ* for a fixed activity *K*. For large *K*, the generating function takes on of large deviations (*LD*) form: $Z_K(x) \sim e^{Kg(x)}$.

In work [60], the quantity corresponding to $Z_K(\gamma)$, the Laplace transform of the First Passage Time (*FPT*) distribution, acts as a non-equilibrium partition function. In work [62], it is shown that $\rho_+(s) \to -g(s),\ \ g(s) = \theta^{-1}(s)$, as noted in work [79]. Here, *K* represents the dynamical activity, $Z_\tau(s) \sim e^{\tau\theta(s)}$, *τ* is the total time, and $Z_\tau(s)$ is the *MGF* of *K(t)*. The function *g* is the functional inverse of *θ* and vice versa (works [79]), given by: $\theta(s) = g^{-1}(s),\quad g(\gamma) = \theta^{-1}(\gamma)$, $s = g(\gamma),\quad \gamma = \theta(s)$. In order not to repeat the designation *s* from the formula $Z_\tau(s) \sim e^{\tau\theta(s)}$, we rewrite equation (8) as $k(r) = \gamma$. Comparing expressions Eqs. (7)-(8) with the expression $Z_\tau(s) \equiv \sum_K e^{-sK} P_\tau(K) = \boldsymbol{E} e^{-sK}$, we obtain that $k(r=-s) = \theta(s)$. The inverse function $k^{-1}(r) = r(\gamma)$ corresponds to the functions $\rho_\pm(\gamma)$. When $\rho_\pm(\gamma) = -s$, $k(r = \rho_\pm(\gamma)) = \theta(s)$. In this case, the dependencies shown in Figs. *1a, b*, pass into the dependencies obtained in the thermodynamics of trajectories (Refs. [74–79]) for the active phase at *s*<0. The function $\rho_-(\gamma = 0) = 0$, as $\theta(s=0) = 0$. Below are the dependencies $\rho_\pm(\gamma)$ for the process of Brownian motion and for the Erlang and Pareto distributions of the fluctuation magnitude.

The *x*-axis in Figs.1 a, b shows the parameter $s=s_{tr}$, where we have designated the parameter $s=s_{tr}$ from the theory of risk. According to the Lundberg equation (8), $s_{tr}=k(r)$. For $r=-s_{tt}$, where $s=s_{tt}$ is the parameter *s* from the thermodynamics of trajectories, $k(r=-s_{tt})=\theta(s_{tt})$. Thus, when replacing $r=-s_{tt}$, the parameter $s=s_{tr}$ along the *x*-axis is $\theta(s_{tt})$. This value goes over to the *y*-axis in the new figure. The value $r=\rho_\pm(s_{tr})$, where $\rho_\pm(s=s_{tr})$ is the *y*-axis variable in Fig. 1a and 1b. We have replaced $r = \rho_\pm(s = s_{tr}) = -s_{tt}$. The index *tr* denotes membership in the set $r \in (s_{tr})$ - in risk theory; index *tt* in $s_{tt}$ - in the thermodynamics of trajectories (different *s* in risk theory and in the thermodynamics of trajectories). This *y*-axis goes over to the *x*-axis in the new figure. The *x*- and *y*-axes swap places. In this case, Fig.1b transforms into Fig.1d at $s=s_{tt}$, known in thermodynamics of trajectories as the behavior of the analogue of the dynamic free energy *θ(s)* per unit time from *s* in the active phase [74-79]. Fig.1b "turns" to the left, and the dependence $\rho_-(s)$ on $s=s_{tr}>0$ transforms into the dependence *θ(s)* on $-s=-s_{tt}<0$ in the active phase of thermodynamics of trajectories. Fig.1b for is chosen because at *m*<0 $\rho_-(s)_{s\to 0} \to 0$, as $\theta(s=0) = 0$. At *m*>0 the situation changes, and $\rho_+(s)_{s\to 0} \to 0$. Such connections between risk theory and thermodynamics of trajectories have not been established before.

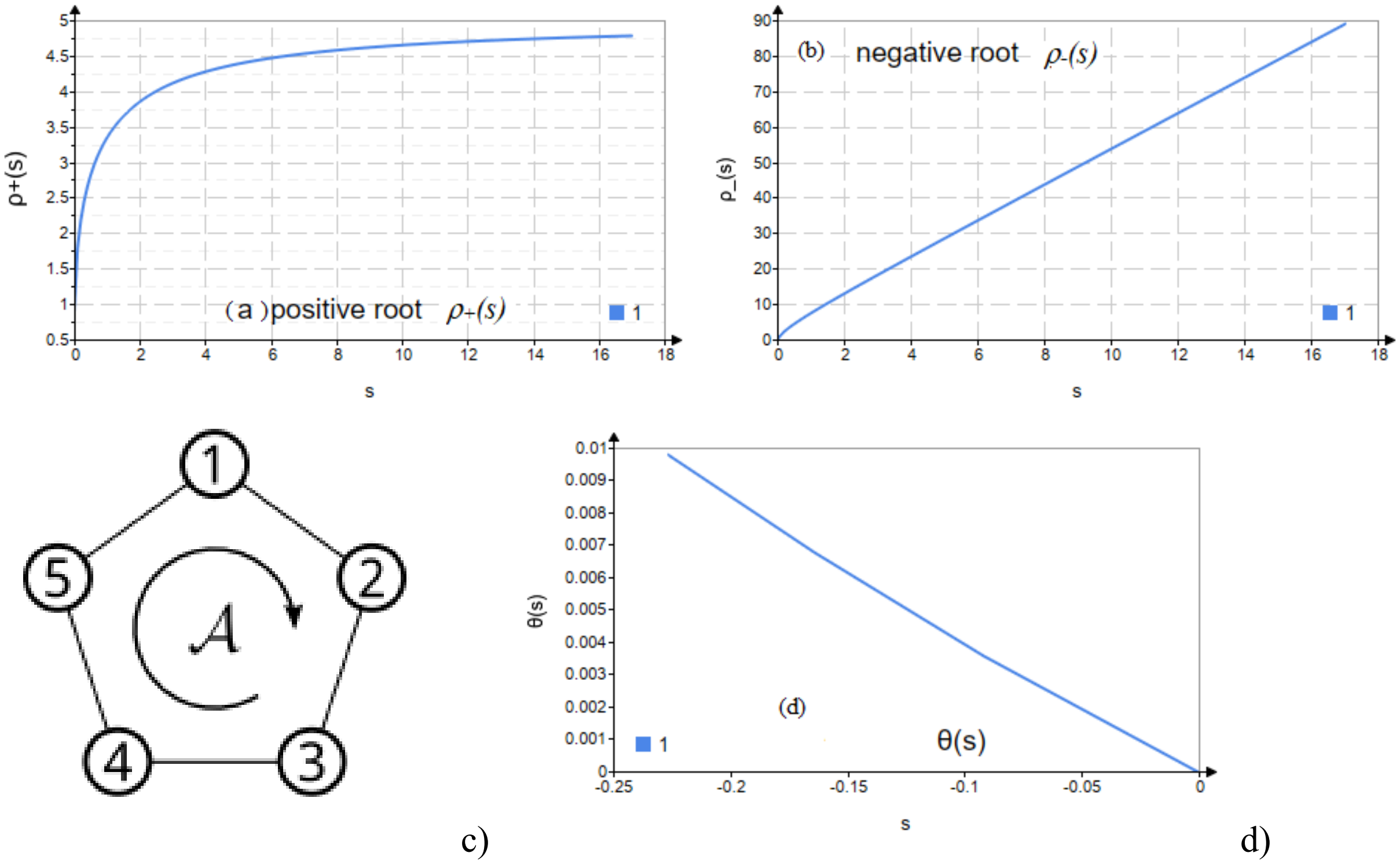

Fig1. Fig. 1a. Positive root Eq. (28) of Lundberg equation (14) with *SCGF* Eq. (26). Fig. 1b. Negative root Eq. (28) of Lundberg equation (14) with *SCGF* Eq. (26). Fig. 1c. (1(a), work [69]) Unicyclic network with affinity *A* and five states, *N*=5. Fig.1d. The analogue of the dynamical free energy per unit time *θ(s)* from *s* in the active phase obtaining from Fig.1b.

### 4.2. First-passage time

Let us now consider the First-Passage Time (*FPT*), Eqs. (5)-(6) for the process *ζ(t)* using the expressions (26)-(28) derived above. The first-passage time is the number of cycles, which may be fractional, expressed as $l+k/N$, where $0<k<N$, $N$ is the number of cycle vertices, and $l$ is the number of full cycles completed, as shown in Fig. 1c (Fig. 1a from Ref. [69]).

Next, we consider expression (17). Differentiating Eq. (17) with respect to *s* and setting *s*=0, we obtain the following dependence:

$$\boldsymbol{E}[\tau^{+}(u),\,\tau^{+}(u)<\infty]=\frac{\lambda}{c}\Big[\frac{\partial\rho_{-}(s)}{\partial s}\frac{1}{\rho_{-}(s)+b}+u\frac{\partial\rho_{+}(s)}{\partial s}\Big]\frac{e^{-\rho_{+}(s)u}}{\rho_{-}(s)+b}\Big|_{s=0}\,. \tag{29}$$

This dependence on $u$ at $s$=0 is shown in Fig. 2a, where $\tau(u)=\inf\{t:R(t)<0\}$, $\tau^{+}(u)=\inf\{t:\zeta(t)>u\}$ (from Eq. (6)) represents the bankruptcy moment, i.e., the time when the process *R(t)* reaches level 0, and the process *ζ(t)* reaches level *u*.

Following the approach in works [62, 80], we differentiate expression (17) with respect to *s*, but without setting *s*=0. The parameter *s* can be considered as a physical field conjugate to the random variable $\tau^{+}$. As a result, we obtain a dependence on two arguments, *u* and *s*, which takes the form:

$$-\frac{\partial\ln\boldsymbol{E}[e^{-s\tau^{+}(u)},\,\tau^{+}(u)<\infty]}{\partial s}=\frac{\boldsymbol{E}[\tau^{+}(u)e^{-s\tau^{+}(u)},\,\tau^{+}(u)<\infty]}{E[e^{-s\tau^{+}(u)},\,\tau^{+}(u)<\infty]}=\frac{\partial\rho_{-}(s)}{\partial s}\frac{1}{\rho_{-}(s)+b}+u\frac{\partial\rho_{+}(s)}{\partial s}, \tag{30}$$

$$u=0,\ \ m<0,\ \ q_+(s)=\boldsymbol{E}[e^{-s\tau^+(0)},\tau^+(0)<\infty]=\frac{\lambda}{c}\int_0^\infty e^{-z\rho_-(s)}\bar{F}(z)dz=\frac{\lambda}{c}\frac{1}{\rho_-(s)+b},$$

$$\boldsymbol{E}[e^{-s\tau^+(u)},\ \tau^+(u)<\infty]=\frac{\lambda}{c}\frac{1}{\rho_-(s)+b}e^{-\rho_+(s)u},$$

$$\boldsymbol{E}[\ \tau^+(u)<\infty]=\boldsymbol{E}[e^{-s\tau^+(u)},\ \tau^+(u)<\infty]\big|_{s=0}=\frac{\lambda}{c}\frac{1}{b}e^{-\rho_+u},\quad \rho_-(s)_{|s=0}=0,\quad \rho_+(s)_{|s=0}=bp_+>0.$$

- Fig. 2a shows the resulting dependence of expression (29) on *u*, for a fixed value of *s*=0.
- Fig. 2b illustrates the dependence of expression (30) on *s*, for a fixed value of *u*=4.
- Fig. 2c presents the dependence of expression (30) on *u*, for a fixed value of *s*=0.

In both cases (Fig. 2b and 2c), for negative values of *s*, the values $\boldsymbol{E}[\tau^+(u),\ \tau^+(u)<\infty]$ become either imaginary or negative. The magnitude $\tau^+(u)$ in expression (17) takes on positive values. The characteristic function Eq. (17) transitions into the Laplace transform. For negative values of the parameter *s* that exceed a certain small value, the Laplace transform diverges, resulting in the values $\boldsymbol{E}[\tau^+(u),\ \tau^+(u)<\infty]$ becoming either imaginary or negative.

However, Figs. 2a and 2c differ significantly, not only for *s*=0. This discrepancy arises because the conditional probabilities in expressions (17) and (29) contain the probability that $\boldsymbol{E}[\ \tau^+(u)<\infty]$ (the bankruptcy moment) occurs. The average value $-\partial\boldsymbol{E}[e^{-s\tau^+(u)},\ \tau^+(u)<\infty]/\partial s\big|_{s=0}$ is influenced by two factors: $\boldsymbol{E}[\ \tau^+(u)<\infty]$ and the average value $-\partial\boldsymbol{E}[e^{-s\tau^+(u)}]/\partial s\big|_{s=0}$.

A more accurate definition in this context is provided by expression (30), as the process $\zeta(t)=ct-S(t)$ does not depend on *u*. As noted in the conclusion of [80], the behavior of $\boldsymbol{E}[\tau^+(u)e^{-s\tau^+(u)},\ \tau^+(u)<\infty]/\boldsymbol{E}[e^{-s\tau^+(u)},\ \tau^+(u)<\infty]$ matches what is illustrated in Fig. 2c. In expression (30), the value $\boldsymbol{E}[\tau^+(u)]$ is reduced accordingly.

The physical interpretation of $\rho_\pm$ in the thermodynamics of trajectories is given in [62, 80] and noted after equation (28): the positive root of equations (8), (14), $\rho_+$, corresponds to the function $-g(s)$, where $g(s)$ is proportional to the logarithm of the partition function (*MGF*). This partition function represents the *MGF* for random time *τ*. If $\rho_+$ corresponds to the function $-g(s)$, represents the *MGF* for random time $\tau^+$, then the function $\rho_-$ represents the *MGF* for random time $\tau^-$. This can be seen from a comparison of the expressions for $\tau^+(u)$ (17) and $\tau^-(u)$ the form $\boldsymbol{E}[e^{-s\tau^-(u)},\ \tau^-(u)<\infty]=q_-(s)e^{\rho_-(s)u},\ u<0$. In thermodynamics of trajectories, the case *u*<0 and the corresponding *MGF* are not considered. It is also possible to consider the physical meaning of the quantity $\rho_-$ from a comparison with the thermodynamics of trajectories, since the quantity $\rho_-$ at *m*<0 is greater than the quantity $\rho_+$ corresponding to the analogue of the dynamical free energy per unit time, the quantity $\theta(s)$.

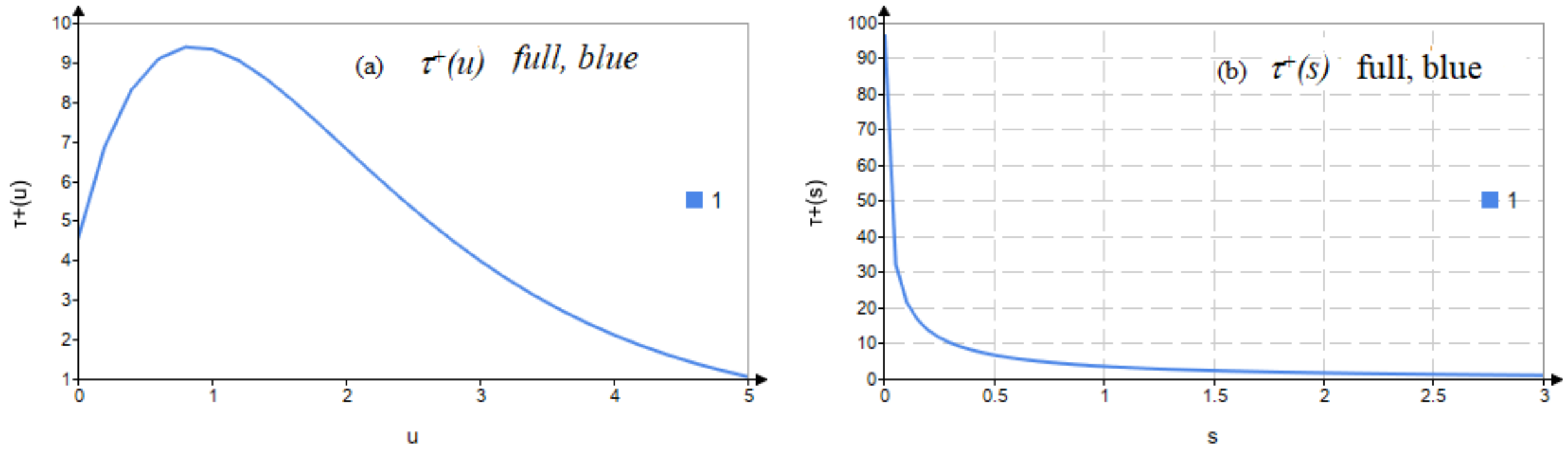

Fig.2a. Dependence of the average value $\boldsymbol{E}[\tau^{+}(u), \tau^{+}(u) < \infty]$ Eq. (29) on $u$, in the interval $u$=0,…, 5 at $s$=0.

Fig.2b. Dependence of the average value $\boldsymbol{E}[\tau^{+}(u) e^{-s\tau^{+}(u)}, \tau^{+}(u) < \infty] / \boldsymbol{E}[e^{-s\tau^{+}(u)}, \tau^{+}(u) < \infty]$ Eq. (30) on $s$, in the interval $s$=0.1,…, 3 at a fixed value $u$=4.

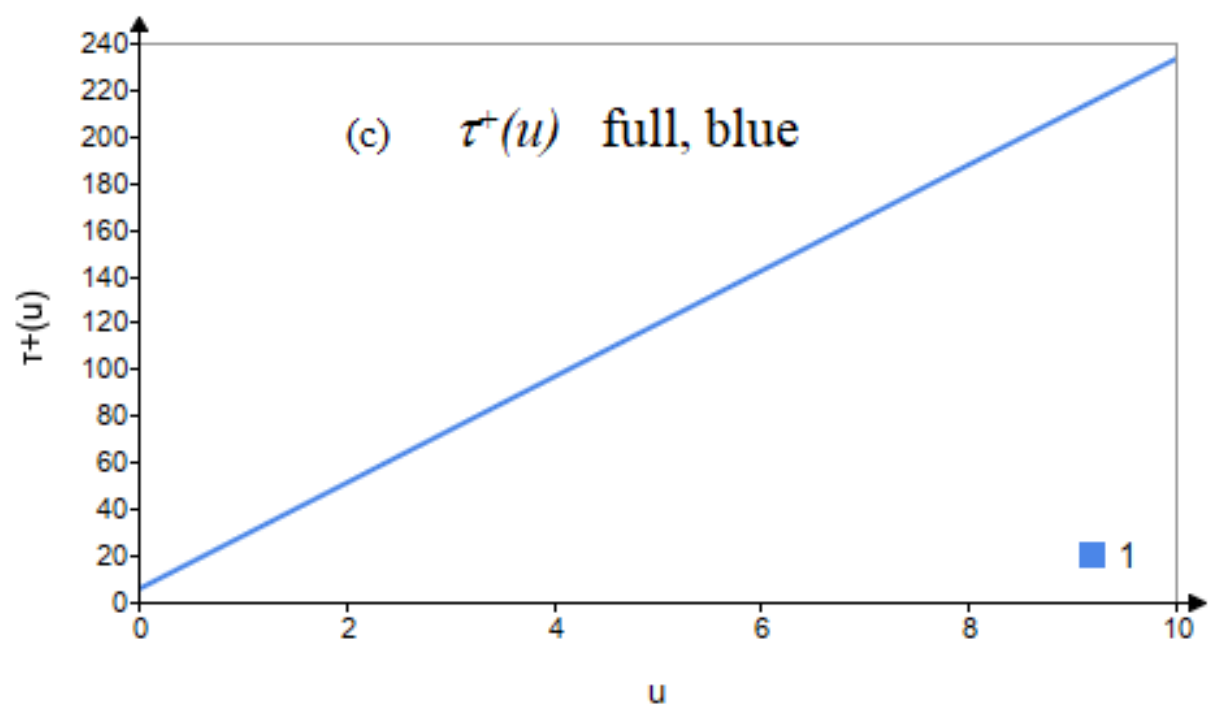

Fig.2c. Dependence of the average value $\boldsymbol{E}[\tau^{+}(u) e^{-s\tau^{+}(u)}, \tau^{+}(u) < \infty] / \boldsymbol{E}[e^{-s\tau^{+}(u)}, \tau^{+}(u) < \infty]$ Eq. (30) on $u$, in the interval $u$=0,…, 10 at a fixed value $s$=0.

Similar relationships can be written for the first-passage time (*FPT*) when reaching negative levels, $\tau^{-}$.

### 4.3. Extremums of functions $\zeta^{\pm}(\theta_s)$

For the extremums of the function $\zeta^{\pm}(\theta_s)$ given in Eq. (15), we examine the dependence not on time *t*, but on an exponentially distributed random variable $\theta_s$. For the characteristic function of the process $\xi(\theta_s)$, defined as $\boldsymbol{E}e^{i\alpha\xi(\theta_s)} = s\int_{-\infty}^{\infty} \boldsymbol{E}e^{i\alpha\xi(t)} e^{-st} dt$, Eq. (0.5), the following relation is written:

$$\varphi(s,\alpha) := \boldsymbol{E}e^{i\alpha\xi(\theta_s)} = \frac{s}{s - \Psi(\alpha)},$$

where the characteristic function of a homogeneous process $\xi(t),\ t \ge 0$ is determined in the theory of random processes, works [7, 9-10, 12-14], by the relations Eqs. (7) and (13).

Let us consider the extremums of the process $\zeta(\theta_s)$, which in our example describes the movement along the cycle (Fig. 1c) of the process given by equations (2) or (4). The *MGF* of the extrema $\zeta^{\pm}(\theta_s) = \sup(inf)_{0 \le t' \le \theta_s} \zeta(t')$ is given by equation (15):

$$\boldsymbol{E}e^{-z\zeta^{+}(\theta_s)} = \frac{p_{+}(s)(b+z)}{\rho_{+}(s)+z}, \quad \rho_{+}(s) = bp_{+}(s),$$

$$\boldsymbol{E}\zeta^{+}(\theta_s) = -\frac{\partial \boldsymbol{E}e^{-z\zeta^{+}(\theta_s)}}{\partial z}_{|z=0} = \frac{b-\rho_{+}(s)}{b\rho_{+}(s)}. \tag{31}$$

If we define the average as:

$$\boldsymbol{E}[\zeta^{+}(\theta_s,\, z)] = -\frac{\partial \ln \boldsymbol{E}[e^{-z\zeta^{+}(\theta_s)}]}{\partial z} = \frac{\boldsymbol{E}[\zeta^{+}(\theta_s)e^{-z\zeta^{+}(\theta_s)}]}{\boldsymbol{E}e^{-z\zeta^{+}(\theta_s)}} = \frac{b-\rho_{+}(s)}{(b+z)(\rho_{+}(s)+z)}, \tag{32}$$

then, for $z$=0, equation (32) coincides with equation (31). However, the parameter z can be interpreted as a field conjugate to the variable $\zeta^{+}(\theta_s)$.

At

$\rho_{+}(s) = [-(s-0.18)+\sqrt{s^2+3.638s+0.0324}]/0.4, \quad \rho_{-}(s) = [(s-0.18)+\sqrt{s^2+3.638s+0.0324}]/0.4$, Eqs. (27) and (28), at $b$=$N$=5, from Eq. (32) we have:

$$\boldsymbol{E}[\zeta^{+}(\theta_s,\, z)] = \frac{\boldsymbol{E}[\zeta^{+}(\theta_s)e^{-z\zeta^{+}(\theta_s)}]}{\boldsymbol{E}e^{-z\zeta^{+}(\theta_s)}} = \frac{5-[-(s-0.18)+\sqrt{s^2+3.6385+0.0324}]/0.4}{(5+z)([-(s-0.18)+\sqrt{s^2+3.6385+0.0324}]/0.4+z)}. \tag{33}$$

Here, $s$ is a field conjugated to time $t$ (or $\theta_s$), representing some external influences that affect the speed of the process $\zeta(t)$, while $z$ is a field conjugated to the extremum $\zeta^{+}(\theta_s)$.

For the minimum values of the function on the interval defined by equation (15), we can evaluate the extremums in a similar manner:

$$\boldsymbol{E}e^{-z\zeta^{-}(\theta_s)} = \frac{\rho_{-}(s)[1-\rho_{-}(s)/b]}{\rho_{-}(s)+z}, \quad cp_{+}(s)\rho_{-}(s) = s,$$

$$\boldsymbol{E}\zeta^{-}(\theta_s) = -\frac{\partial \boldsymbol{E}e^{-z\zeta^{-}(\theta_s)}}{\partial z}\bigg|_{z=0} = \frac{1-\rho_{-}(s)/b}{\rho_{-}(s)}, \tag{34}$$

$$\boldsymbol{E}[\zeta^{-}(\theta_s,\, z)] = -\frac{\partial \ln \boldsymbol{E}[e^{-z\zeta^{-}(\theta_s)}]}{\partial z} = \frac{\boldsymbol{E}[\zeta^{-}(\theta_s)e^{-z\zeta^{-}(\theta_s)}]}{\boldsymbol{E}e^{-z\zeta^{-}(\theta_s)}} = \frac{1}{(\rho_{-}(s)+z)}. \tag{35}$$

For $z$=0, equation (35) does not coincide with equation (34).

The dependencies of equations (33) (red) and (34) (green) on $s$ at $z$=0 is shown in Fig. 3. As $z\to\infty$, the function behaves as $\boldsymbol{E}\zeta^{\pm}(\theta_s, z) \to 0$.

The quantity $\zeta^{+}(\theta_s)$ is distributed according to the probability $\bar{P}_{+}(s,x) = q_{+}(s)e^{-\rho_{+}(s)x}$, $x>0$. The probability of bankruptcy, that is, the process $R(t)$ reaching negative values, or the process $\zeta(t)$ reaching a level $u$>0, is determined by the "tail" of the distribution $\zeta^{+}$, $\Psi(u) = P\{\zeta^{+} > u\} = \bar{P}_{+}(u)$. For the model in equation (12), $\Psi(u) = q_{+}e^{-\rho_{+}u}$, $q_{+} = \lambda / cb$.

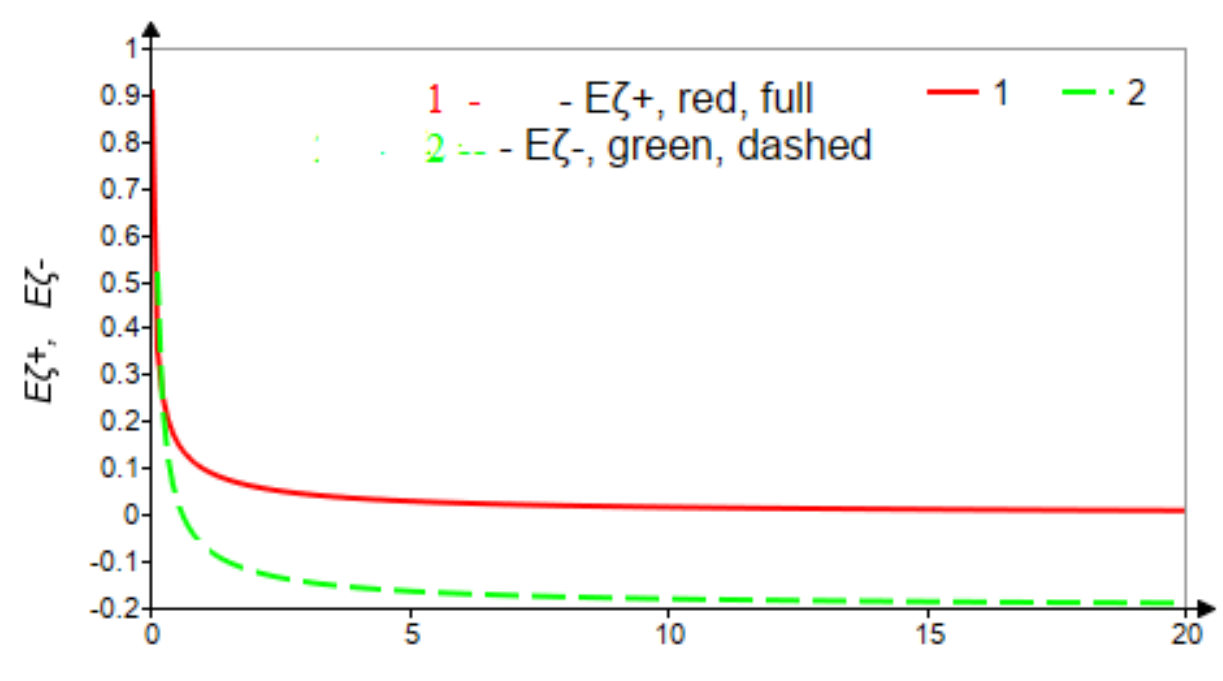

Fig.3. Dependencies Eq. (33) $\boldsymbol{E}\zeta^{+} = \boldsymbol{E}\zeta^{+}(\theta_s, z)$ (red, full) and Eq. (34) $\boldsymbol{E}\zeta^{-} = \boldsymbol{E}\zeta^{-}(\theta_s, z)$ (green, dash) from *s* at *z*=0.

### 4.4. Moments of return

For the moment $\tau'(u) = \inf\{t > \tau(u), R(t) > 0\}$, which is the moment of return of the process $R(t)$ after bankruptcy to the half-plane $\Pi^{+} = \{y > 0\}$, the *MGF* is given by equation (17):

$$\varphi_u(s) = \boldsymbol{E}[e^{-s\tau'(u)}, \tau^{+}(u) < \infty] = q_{+}(s,u)\frac{b}{b+\rho_{-}(s)}, \tag{36}$$

where $q_{+}(s,u) = \frac{\lambda}{c}\frac{1}{b+\rho_{-}(s)}e^{-\rho_{+}(s)u}$.

In the process under consideration, a return is possible when the direction of movement changes. In some problems, it is important to know the moment when the process $\zeta(t)$ returns from a position exceeding the level *u* to states below *u*, where *u* corresponds to a parameter of the form *l+k/N*, with 0<*k*<*N*, *N* being the number of vertices in the cycle, and *l* representing the number of cycles passed (Fig. 1c).

The average value of $\tau'(u)$ from Eq. (36) at $s \neq 0$ is equal to:

$$\boldsymbol{E}\tau'(u,s) = -\frac{\partial \ln \varphi_u(s)}{\partial s} = \frac{\boldsymbol{E}[\tau'(u)e^{-s\tau'(u)}, \tau^{+}(u) < \infty]}{\boldsymbol{E}[e^{-s\tau'(u)}, \tau^{+}(u) < \infty]} = \frac{\partial \rho_{+}(s)}{\partial s}u + \frac{\partial \rho_{-}(s)}{\partial s}\frac{2}{b+\rho_{-}(s)}. \tag{37}$$

Figures 4a and 4b show the dependence of the average value of $\tau'(u,s)$, Eq. (37) on *s* for *u*=5, and on *u* for *s*=0, calculated using the parameters corresponding to expressions (28).

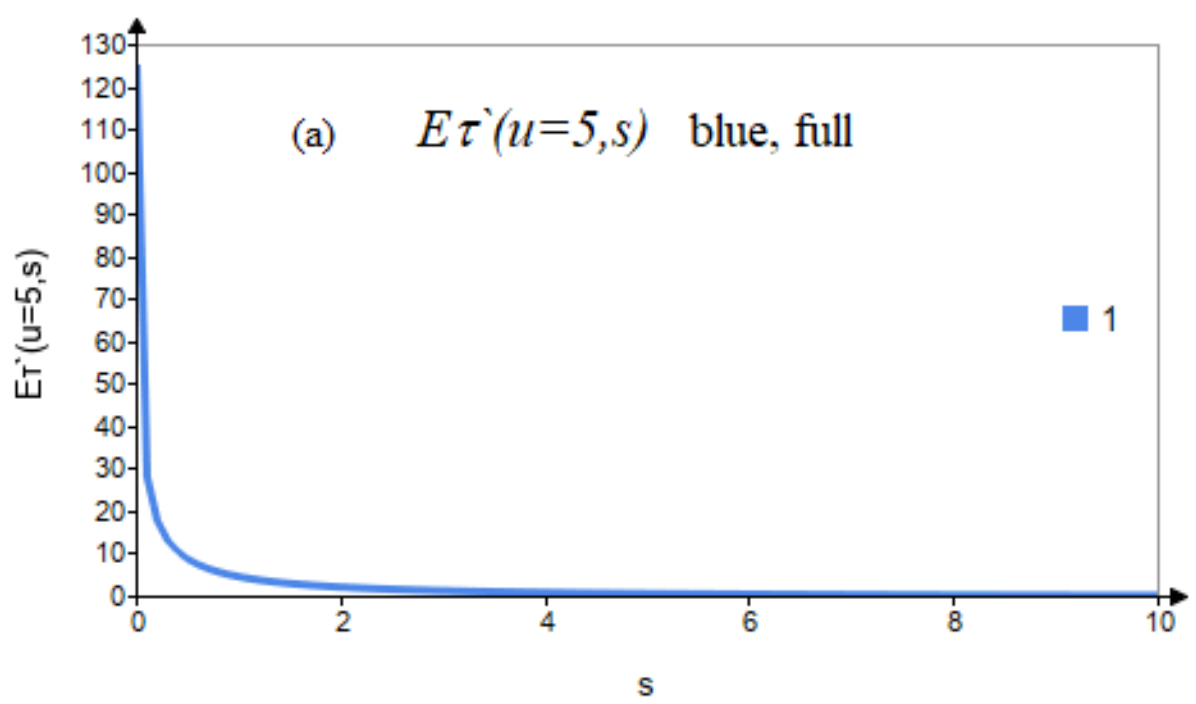

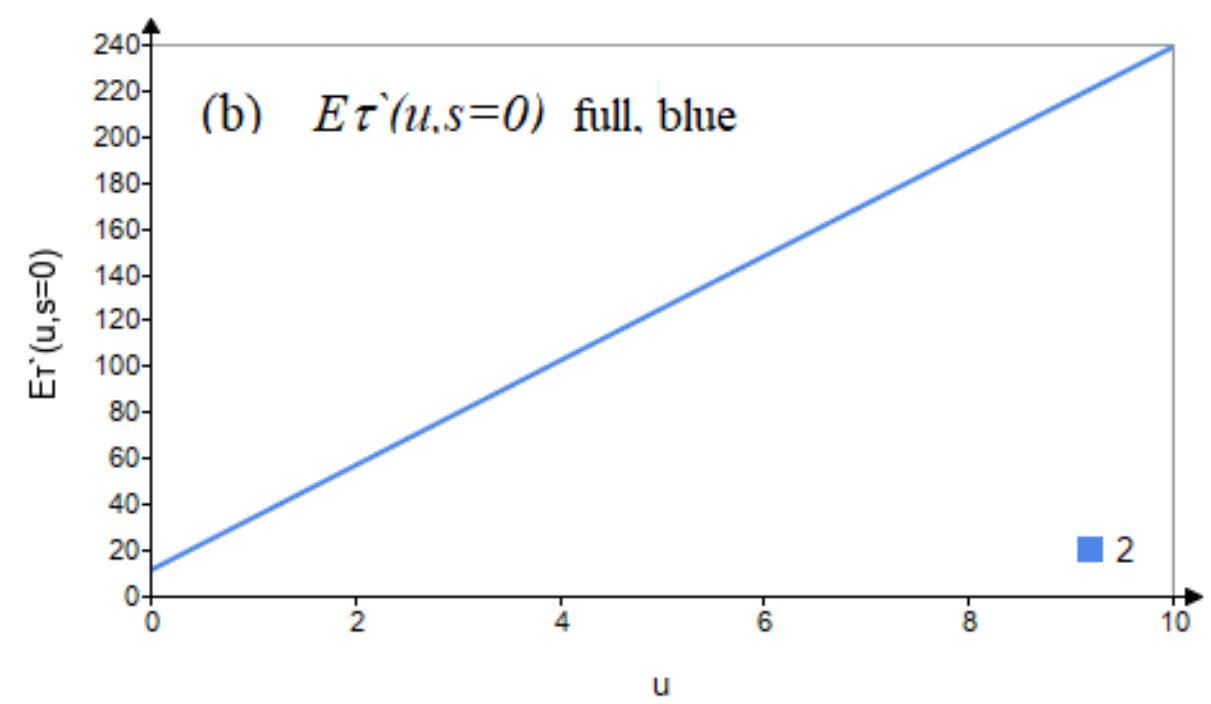

Fig. 4a. Dependence of the mean value $\boldsymbol{E}\tau'(u,s)$ on *s* in the interval *s*=(0,…, 10) at $s \neq 0$, *u*=5.

Fig. 4b. Dependence of the mean value $\boldsymbol{E}\tau'(u,s)$ on *u* in the interval *u*=(0,…, 20) at *s*=0.

It is evident that $\boldsymbol{E}\tau'(u,s)$ takes its maximum values at *s*=0. The effect of the field *s*, conjugate to $\tau'(u,s)$, reduces the average values. For negative values of *s*, the average values become imaginary or negative, meaning they do not exist in this context.

The value $\tau'(u,s)$ is the positive magnitude and the characteristic function Eq. (36) are replaced by the Laplace transform. For negative values of *s*, the Laplace transform may diverge, as indicated in reference [61], which indeed occurs. In reference [60], a statistical distribution is introduced that includes *FPT* as a random thermodynamic parameter and its conjugate thermodynamic parameter (denoted here as *s*). Reference [95] demonstrates that this parameter is proportional either to entropy production or to flows within the system, depending on the context. This parameter equals zero only in the equilibrium state. In general, the relationship between parameter *s* and changes in entropy is more complex. For model systems (two-level, three-level, and others), this relationship is discussed in reference [61]. For charge transport, see reference [65]. These issues have also been addressed in reference [62].

### 4.5. The surplus prior to ruin

Of interest in various physical, biological, and other applications may be the value of the process *R(t)* before bankruptcy, reaching the boundary (zero value for the process *R(t)* and the boundary *u* for the process $\zeta(t)$). In Equation (9), this value is denoted as $\gamma_+(u)$, the value before bankruptcy, the surplus before ruin. In our example, this is the value of the process $\zeta(t)$ (Eq. (2)) before the jump that reaches the value *u*.

This quantity, denoted as $\gamma_+(u)$, represents the value of the process right before it hits the boundary - zero for the process $R(t)$ or *u* for the process $\zeta(t)$. This quantity reflects the state of the system just before reaching critical thresholds.

The *MGF* of this value, $\gamma_+(u)$, Eq. (17), for model Eq. (12) can be written as:

$$\tilde{q}_+(z,u) = \boldsymbol{E}[e^{-z\gamma_+(u)}, \tau^+(u) < \infty] = q_+ e^{-\rho_+ u}\frac{b}{b+z}, \ \ \rho_+ = bp_+ = \frac{bc-\lambda}{c}, \ \ q_+ = \frac{\lambda}{cb}. \tag{38}$$

This *MGF* helps quantify the distribution of the surplus prior to ruin, providing insights into how the system behaves just before failure.

The mean value for *MGF* Eq. (38) is:

$$\boldsymbol{E}[\gamma_+(u)e^{-z\gamma_+(u)}, \tau^+(u) < \infty] = -\frac{\partial \boldsymbol{E}[e^{-z\gamma_+(u)}, \tau^+(u) < \infty]}{\partial z} = \frac{\lambda}{c}\frac{1}{(b+z)^2}e^{-(b-\lambda/c)u}\Big|_{z=0} = \frac{\lambda}{cb^2}e^{-(b-\lambda/c)u}. \tag{39}$$

The behavior of the quantity Eq. (39) depending on *u* at *z*=0 with parameters corresponding to expression (28) is shown in Fig. 5a.

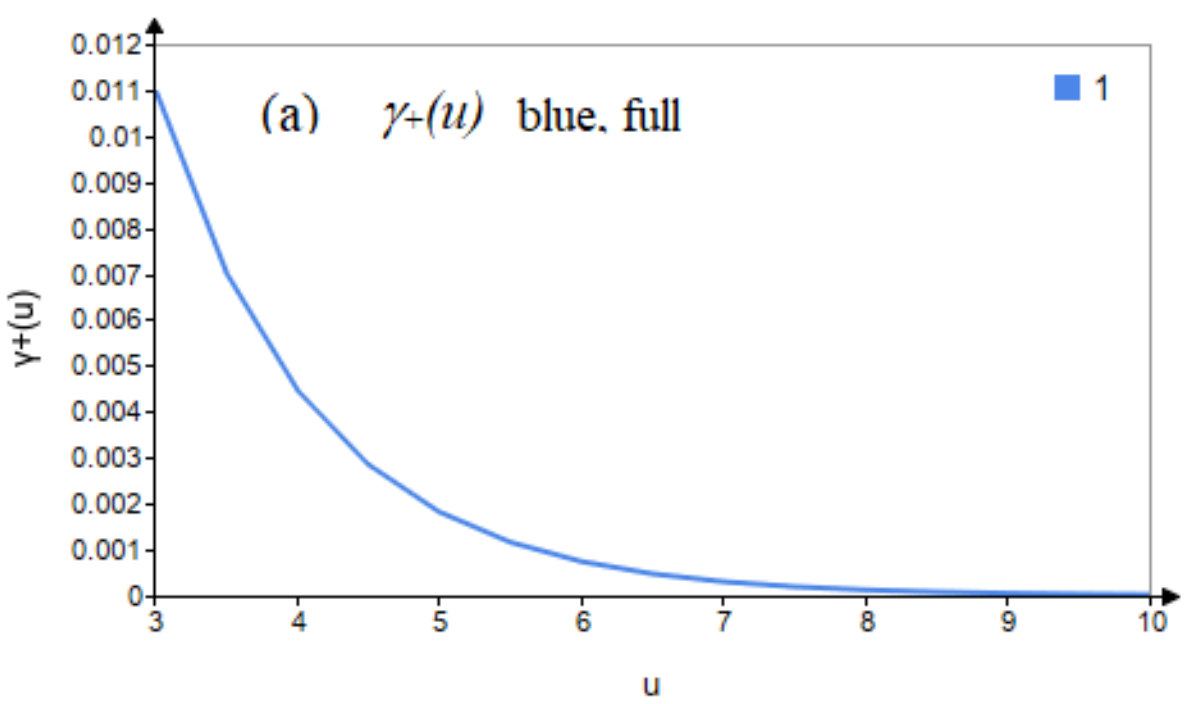

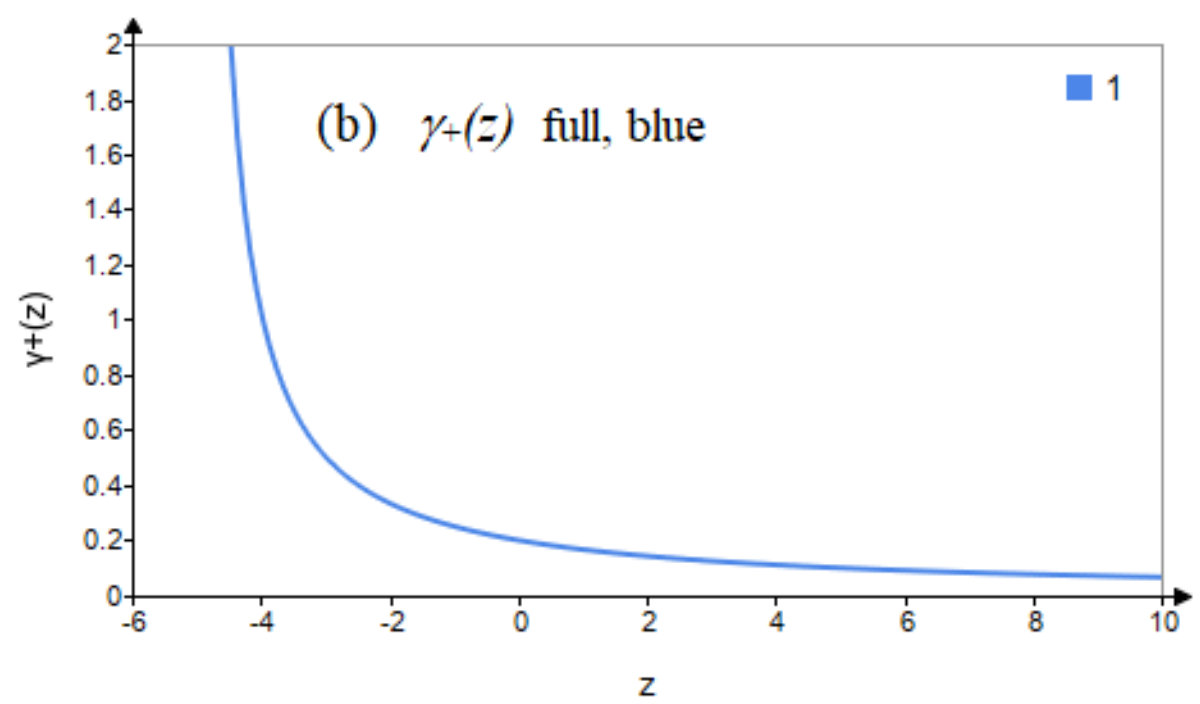

Fig. 5a. Behavior of the mean value Eq. (39) of the quantity $\gamma_+(u)$ depending on *u* at *z*=0 in the interval *u*=3,…, 10. Fig. 5b. Behavior of the mean value Eq. (39) of the quantity $\gamma_+(z)$ depending on *z* at *u*=5, in the interval *z*=-5,…, 10; at $z \neq 0$ in Eq. (39).

The behavior of the quantity Eq. (39) depending on *z*=-5,…, 10 at *u*=5 is shown in Fig. 5b.

The function $\dfrac{\boldsymbol{E}[\gamma_+(u)e^{-z\gamma_+(u)}]}{\boldsymbol{E}[e^{-z\gamma_+(u)}]} = -\dfrac{\partial \ln \boldsymbol{E}[e^{-z\gamma_+(u)}]}{\partial z} = \dfrac{1}{(b+z)}$ does not depend on *u* (for Eq. (12)).

### 4.6. Distribution of the duration of stay of the process $R(t)$ in the lower half-plane $\Pi^-\{x<0\}$

The duration of the process $R(t)$ staying in the lower half-plane (or, equivalently, the duration of the process *ζ(t)* staying above the level *u*) is referred to as the "red time" [67] or the "red period", Eqs. (10), (17). This is denoted as *T`(u)*, and represents the time spent below *u* after the process reaches the threshold and before it returns to the upper half-plane (or before it crosses above *u* again for the process $\zeta(t)$).

The *MGF* for this quantity *T`(u)=inf{t>τ(u), R(t)>0}-τ(u)*, Eq. (10), which is defined as:

$$\boldsymbol{E}[e^{-sT'(u)}, \tau^{+}(u)<\infty] = q_{+}e^{-\rho_{+}u}\frac{b}{b+\rho_{-}(s)}, \tag{40}$$

Eq. (17), where *τ(u)* is the time when the process $\zeta(t)$, Eq. (2), first reaches the threshold *u*.

In this case, the *MGF* helps to analyze the statistical properties of the duration of the red period, giving insights into the frequency and duration of periods where the process $\zeta(t)$ stays in unfavorable conditions (above *u*, or above a critical level for the process $\zeta(t)$). The explicit form of the *MGF* will depend on the particular model used, such as the exponential distribution model described in expression (12).

The average value is:

$$\boldsymbol{E}[T'(u)e^{-sT'(u)}, \tau^{+}(u)<\infty] = \frac{\partial\rho_{-}(s)/\partial s}{(b+\rho_{-}(s))^{2}}\frac{\lambda}{c}e^{-(b-\lambda/c)u}. \tag{41}$$

The dependence of value Eq. (41) on *s* for *u*=5 is shown in Fig. 6a, the dependence on *u* for *s*=0 is shown in Fig. 6b.

The behavior of the mean value Eq. (42) depending on *s* for *s*=(0,…, 10) is shown in Fig. 6c. The result does not depend on *u* (for case Eq. (12)):

$$\boldsymbol{E}[T'(s)] = \frac{\boldsymbol{E}[T'(u)e^{-sT'(u)}, \tau^{+}(u)<\infty]}{\boldsymbol{E}[e^{-sT'(u)}, \tau^{+}(u)<\infty]} = -\frac{\partial \ln \boldsymbol{E}[e^{-sT'(u)}, \tau^{+}(u)<\infty]}{\partial s} = \frac{\partial\rho_{-}(s)/\partial s}{b+\rho_{-}(s)}. \tag{42}$$

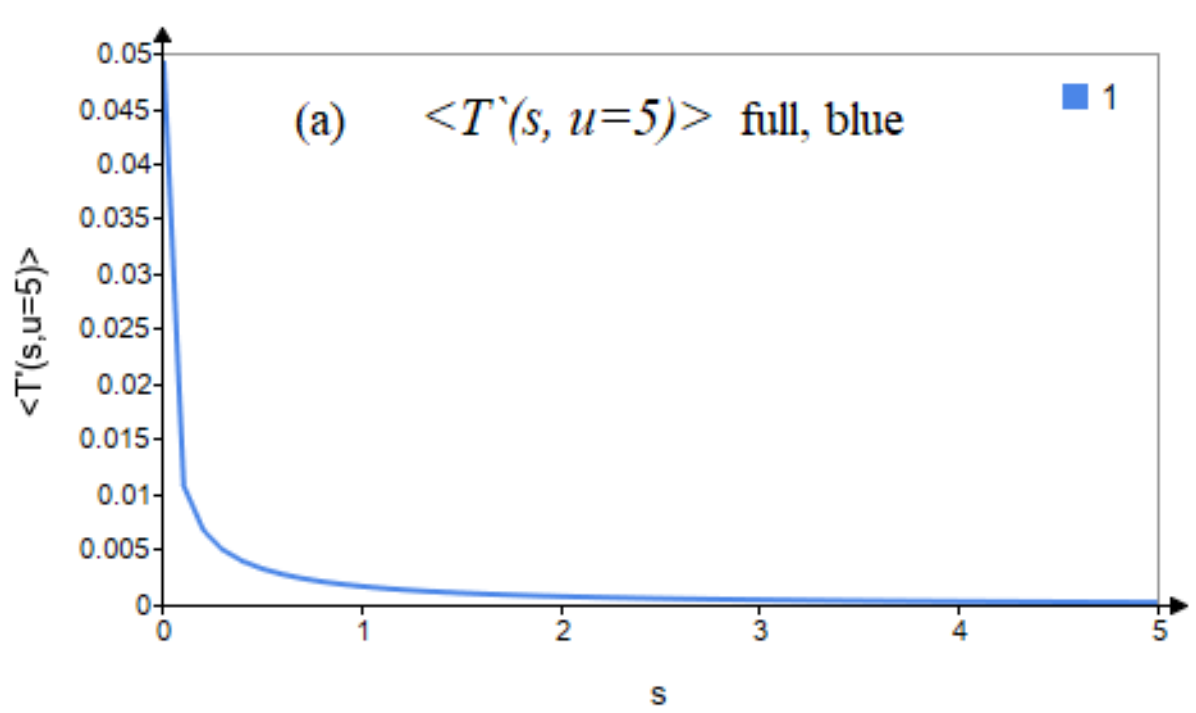

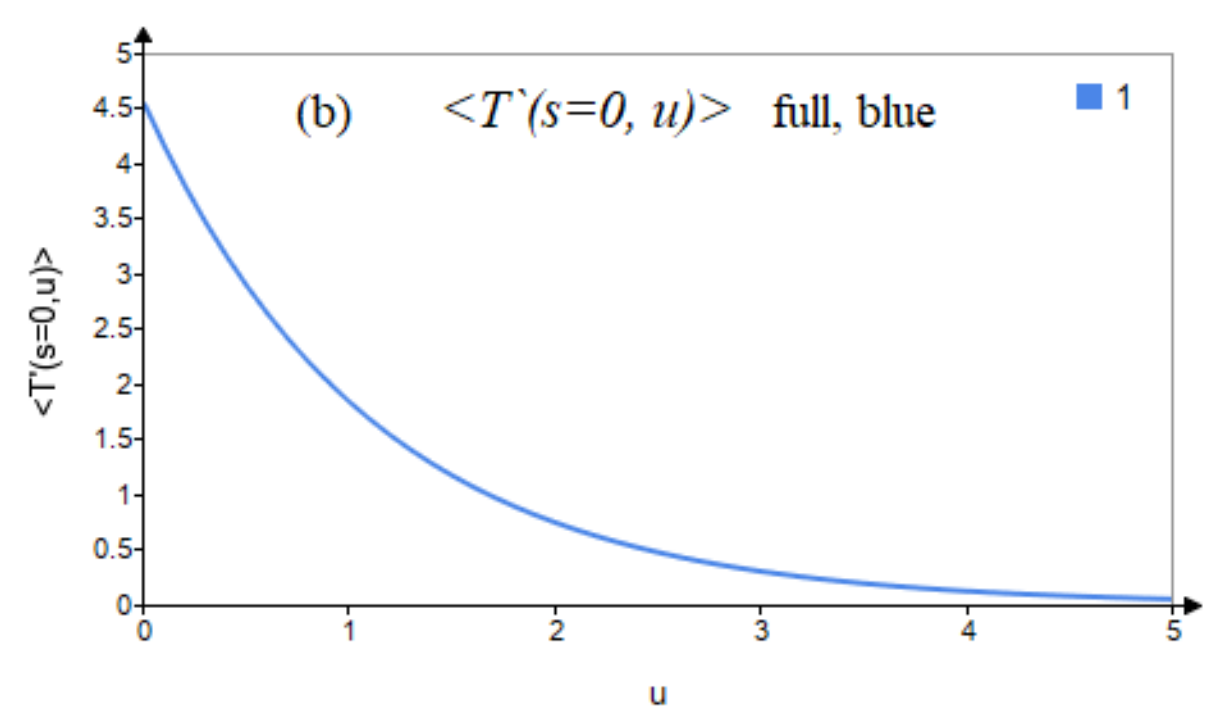

Fig. 6a. Behavior of the mean value $\boldsymbol{E}[T`(u)e^{-sT'(u)}, \tau^{+}(u)<\infty]$ Eq. (41) depending on *s* for *u*=5; *s*=(0,…, 5).
Fig. 6b. Behavior of the mean value $\boldsymbol{E}[T`(u)e^{-sT'(u)}, \tau^{+}(u)<\infty]$ Eq. (41) depending on *u* for *s*=0; *u*=(0,…, 5).

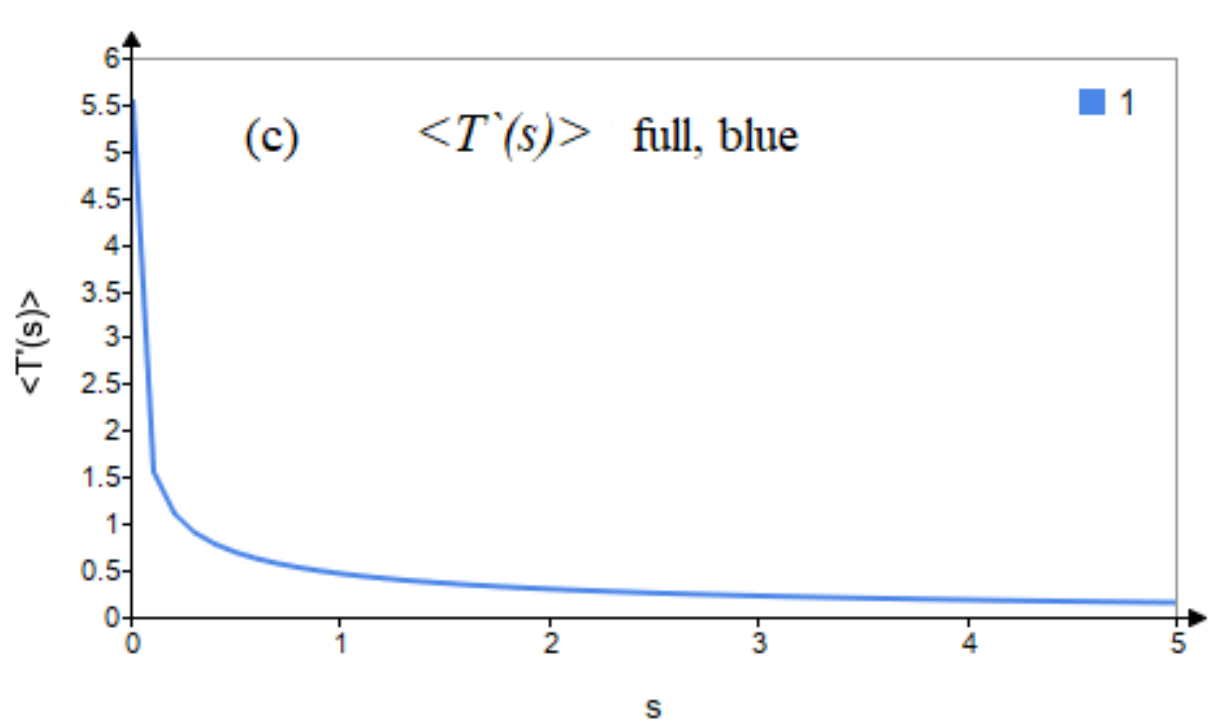

Fig.6c. Behavior of the mean value $\boldsymbol{E}[T'(s)]$ Eq. (42) depending on $s$, $s$=(0,…, 10); does not depend on $u$.

For $s$<0 and $x$<0, $T'(0)$<0, due to the divergence of the partition function. The value Eq. (42), Fig. 6c, does not depend on $u$ for model Eq. (12).

**4.7. Time the process $\zeta(t)$ spends above level $u$.**

Knowledge of the boundary functional $Q_u(\infty)$ of the time the process $\zeta(t)$ spends above level $u$, Eq. (23) may be interesting and useful in various applications.

The time a process $\zeta(t)$ spends above a certain level $u$ can be a crucial boundary functional in various applications, such as in risk theory, queueing theory, or physical, chemical, biological, and other systems. These functional measures the total amount of time the process remains above a critical threshold, which could correspond to system stability or failure prevention in practical situations.

For the model in case Eq. (12), where the jumps in the process are exponentially distributed, the *MGF* of the quantity $Q_u(\infty)$, for the amount of time the process spends above the level $u$ is given by:

$$D_u(\mu) = \boldsymbol{E}e^{-\mu Q_u(\infty)} = 1-(1-\rho_+ / \rho_+(\mu))e^{-\rho_+ u}, \ \ u>0. \qquad (43)$$

The characteristic function $D_u(\mu)$ provides a compact representation of the distribution of this "above-level" time. This expression captures how the time spent above the level $u$ is distributed and provides insights into the frequency and duration of excursions above the critical threshold $u$. This knowledge is particularly important in contexts such as determining the probability of a system remaining stable or exceeding certain performance levels for extended periods, as well as calculating risks or planning for potential system failures.

The average value $Q_u(\infty)$ is:

$$\boldsymbol{E}[Q_u(\infty)] = \frac{\boldsymbol{E}[Q_u(\infty)e^{-\mu Q_u(\infty)}]}{\boldsymbol{E}e^{-\mu Q_u(\infty)}} = -\frac{\partial \ln D_u(\mu)}{\partial \mu} = \frac{(\rho_+ / (\rho_+(\mu))^2)(\partial \rho_+(\mu) / \partial \mu)e^{-\rho_+ u}}{1-(1-\rho_+ / \rho_+(\mu))e^{-\rho_+ u}}. \qquad (44)$$

The behavior of the average $Q_u(\infty)$ Eq. (44) depending on $u$=(0,…, 4) at $s=\mu$=0 is shown in Fig. 7a. The behavior of the mean Eq. (44) depending on $s=\mu$=(1,…, 10) at $u$=0.5 is shown in Fig.7b.

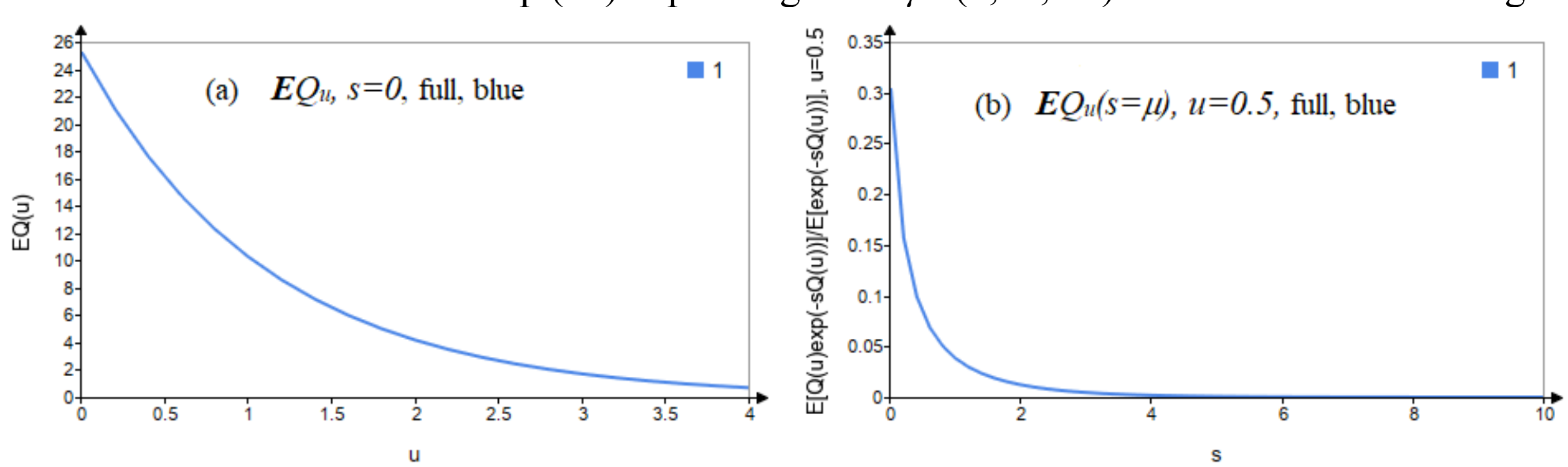

Fig 7a. Behavior of the mean value $Q_u(\infty)$ Eq. (44) depending on $u$=(0,…, 4) at $s=\mu$=0.

Fig. 7b. Behavior of the mean value $Q_u(\infty)$ Eq. (44) depending on $s=\mu$=(1,…, 10) at $u$=0.5.

The functional $Q_u(\infty)$ is very close to the functional $T'(u)$ from subsection 4.6. If we compare Fig. 6a with a similar figure 7c obtained from expression (43), when:

$$\boldsymbol{E}[Q_u(\infty)e^{-\mu Q_u(\infty)}] = -\frac{\partial D_u(\mu)}{\partial \mu} = (\rho_+ / (\rho_+(\mu))^2)(\partial \rho_+(\mu) / \partial \mu)e^{-\rho_+ u}, \qquad (45)$$

then we obtain the result shown in Fig. 7c, which is very close, including quantitative indicators, to Fig. 6a. These two functionals are also close to the functional $\tilde{\sigma}_{N_u}$ defined after Eq. (20) with Eq. (21). After Eq. (23) it is noted that for example Eq. (12) $Q_u(\infty)=\tilde{\sigma}_{N_u}$ . The exponential factors depending on *u* in Eq. (40) and Eqs. (43)-(45) coincide. The pre-exponential factors that do not depend on *u* differ. The advantage of the functional $Q_u(\infty)$ compared to the other two functionals is its generality; it is valid for semi-continuous and almost semi-continuous from above and for arbitrary almost semi-continuous from below risk processes. Therefore, in the next section we will apply it to two other examples.

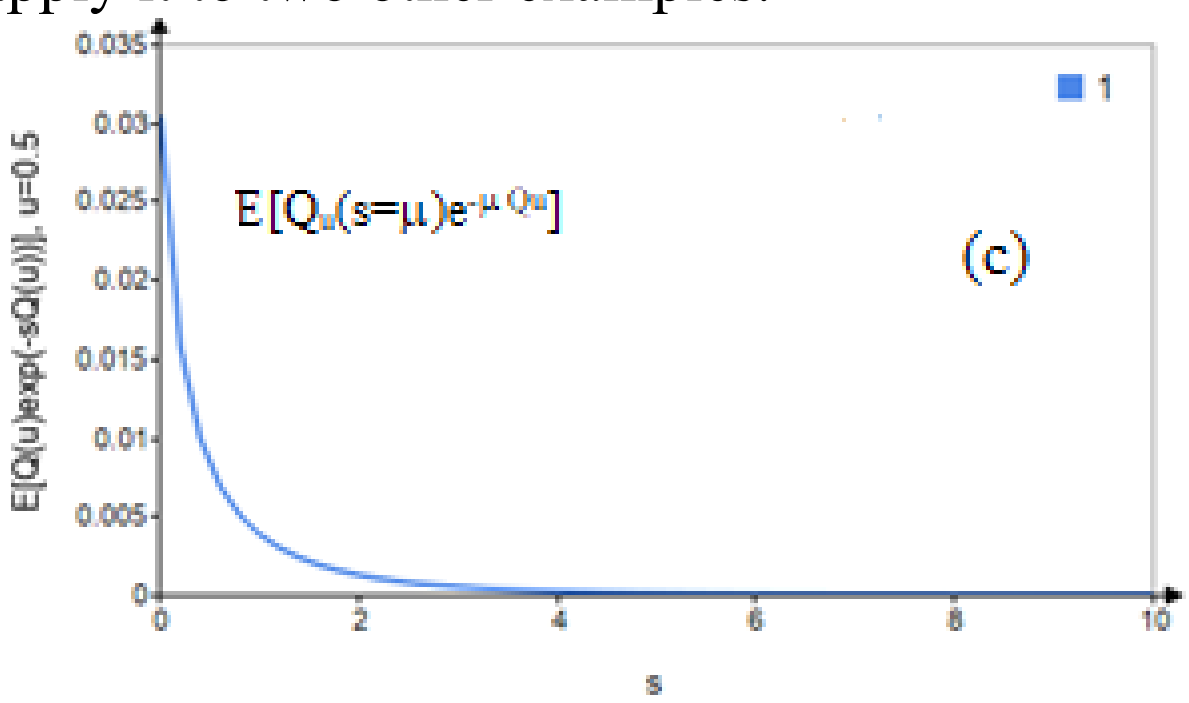

Fig.7c. Behavior of the value $\boldsymbol{E}[Q_u(\infty)e^{-\mu Q_u(\infty)}]$ Eq. (45) from $s=\mu$ in the interval $s=\mu=(0.5,\ldots,10)$ at $u=0.5$.

## 5. Other Examples

In this section, we consider two additional physical scenarios where the process *ζ(t)* spends time above a level *u*, extending the boundary functional example 4.7 to more complex systems. These scenarios include **non-linear diffusion** and **multiple diffusing particles with reversible target-binding kinetics**. Some other examples will be considered: two-level model, Brownian motion. The accuracy of the approximations made in Section 4 is also assessed. Non-exponential heavy-tailed distributions for the magnitude of fluctuations are also considered. For these examples, various boundary functionals are considered (e.g. first-passage time, ruin probability).

### 5.1. First-passage time statistics for non-linear diffusion

In work [81], a nonlinear diffusion equation in which the diffusion coefficient depends on the concentration/probability density as a power law is considered and its fundamental first-passage properties are investigated. To use the expression (44)-(45), we must define $\rho_+(s)$, the positive solution of the Lundberg equation (8) (or Eq. (14) in approximation (12)).

In **non-linear diffusion** processes, the dynamics differ from linear systems due to factors like varying diffusivity depending on the concentration or position of particles. These processes often describe physical phenomena such as heat conduction in inhomogeneous media or the spread of chemicals in non-uniform environments. Let *ζ(t)* be the non-linear diffusion process, where the diffusivity *D[p(x, t)]* depends on the probability density function of finding a diffusing particle in the location *x* at time *t* through which it also depends on the variables *x* and *t*. The boundary functional we are interested in is the amount of time the process spends above level *u*.

For non-linear diffusion processes, the probability density *p(x,t)* evolves according to a generalized diffusion equation:

$$\frac{\partial p(x,t)}{\partial t} = \frac{\partial}{\partial x}(D[p(x,t)]\frac{\partial p(x,t)}{\partial x}),$$

where $D[p(x, t)]$ is a non-constant diffusivity function. The dwell time $Q_u$ above level $u$ in such a system is characterized by the stochastic trajectories of the diffusion process, which are influenced by the varying diffusivity. For non-linear diffusion, the characteristic function (*CF*) of the dwell time $Q_u$ above level $u$ can be written in a form similar to equation (23).

Let's consider the case when the parameter $\sigma$=2, where $\sigma$ is the power-law exponent; the power-law function $D[p(x,t)] = D_0(p(x,t)/p_0)^\sigma$, $p(x,t)$ is a probability density function (*PDF*) of finding a diffusing particle in a location $x$ at time $t$. In this expression $p_0$ denotes a constant reference value of a probability density, whereas $D_0$ is the diffusivity at that reference value. The power-law exponent $\sigma$ is a real parameter that characterizes the nonlinearity appearing in the diffusion term. The number parameter $D[p(x,t)] = D_0(p(x,t)/p_0)^\sigma$ corresponds to the generalized diffusion coefficient of the physical dimension $[D] = L^{\sigma+2}/T$, where $L$ and $T$ are units of the length and the time, respectively.

The final expression for the Laplace transforms of the non-linear diffusion propagator with parameter $\sigma$=2 reads:

$$\tilde{p}(0,s|x') = \frac{a^{3/2}(x)}{\sqrt{2\pi\sqrt{D}}}\exp[-a^2(x)s/2][K_{3/4}(a^2(x)s/2) - K_{1/4}(a^2(x)s/2)], \qquad (46)$$

where $a(x) = p(x-x')^2/4D^{1/2}$, $K_\nu(z)$ is the modified Bessel function of the second kind [81].

The Laplace transformation of the solution of the equation for nonlinear diffusion on the half-axis $x$=(0,∝) has the form $\tilde{p}(0,s|x') = \int_0^\infty p(0,t|x')e^{-st}dt$, where $p(x,t|x'=x_0)$ is the *PDF* of appearing a particle in $x$ at time $t$, if it initially occupied the position $x_0$ at time $t$=0.

Then in the Lundberg equation (8) $k(r) = \ln \tilde{p}(0,r|x')$. From Eqs. (8) and (46) we obtain a complex equation for determining the root $r_s = \rho_+(s) > 0$. It is relatively easy to determine the time spent above level 0: $\boldsymbol{E}e^{-\mu Q_0(t)} = \frac{2}{\pi}\int_0^t e^{-\mu y}\frac{dy}{\sqrt{t^2-y^2}}$. The average time spent above level 0 is determined by the integral $\langle Q_0(t)\rangle = -\frac{\partial \boldsymbol{E}e^{-\mu Q_0(t)}}{\partial\mu}\Big|_{\mu=0} = \frac{2}{\pi}t^2{}_{t\to\infty} \to \infty$ for the process $\zeta(u),\ u\in[0,t]$. This is obvious, since the process is considered on the positive semiaxis.

In work [81] asymptotic formulas for $\sigma$-dependent first-passage time distributions were also obtained, but they also have a complex form. As for the case $\sigma$=2, they can be solved numerically or using various approximations. Section 5.5 discusses Brownian motion, which is closely related to ordinary diffusion.

### 5.2. Multiple diffusing particles with reversible target-binding kinetics

Certain biochemical reactions can only be triggered after binding of a sufficient number of particles to a specific target region such as an enzyme or a protein sensor. Problems of this kind were studied in Refs. [30, 34, 40, 82–88].

In systems of **multiple diffusing particles**, such as those in chemical or biological settings, we may be interested in the time particles spend in a specific region of space or bound to a target.

Reversible binding kinetics means particles can stochastically bind and unbind from targets, introducing additional complexity to the process.

Consider a system with $N$ particles diffusing in space, where each particle can bind to a target with rate $k_{on}$ and unbind with rate $k_{off}$. The process $\zeta(t)$ represents the number of particles bound to the target at time $t$.

The boundary functional of interest is the total time $Q_x$ spent by the number of bound particles above a certain threshold $x$ (i.e., the number of particles bound is greater than $x$) over time $t$. This process can be modeled as a continuous-time Markov process with transitions governed by the binding and unbinding rates.

The concept of the **time spent above a level** can be applied to a wide variety of physical, biological, and chemical systems. By analyzing the characteristic function $\boldsymbol{E}[Q_x(\infty)]$, we gain insights into the distribution of times the process exceeds a given threshold, which can be important in understanding system stability, reaction times, or the likelihood of specific outcomes.

In work [30], the class of **diffusion-controlled reactions** where the reaction is initiated upon the binding of a given number $K$ of $N$ independently diffusing particles to a target region was studied. These types of reactions are particularly relevant to biological and biochemical systems where a specific event, such as cellular signaling or enzymatic reactions, is triggered when a set number of molecules bind to a target site.

In several examples cited in Ref. [30], a biochemical event, such as signaling, is initiated when a fixed number $K$ of $N$ diffusing particles bind to the target region for the first time. If $N(t)$ denotes the number of bound particles at time $t$, the reaction time $T_{K,N}=inf\{t>0: N(t)=K\}$ is the time of first crossing a fixed threshold $K$ by the stochastic non-Markov process $N(t)$. However, in most cases, the binding is reversible, meaning that some particles may detach and resume diffusion before the $K$-th fastest particle binds, making the problem of "impatient" particles [57] much more difficult. The process $N(t)$ can thus be approximated by a Markovian birth-death process.

In work [30], the reaction time $T_{K,N}=inf\{t>0: N(t)=K\}$ is defined as the first-crossing time of a fixed threshold $K$ by the stochastic non-Markovian process $N(t)$. However, this only represents the reaction start time. Additional time is required for the reaction to proceed. The results from subsection 4.7, concerning the time the process $\zeta(t)$ Eq. (2) spends above level $x$, can be applied to address this problem.

In Ref. [30], the relation for diffusion-controlled reactions is given by the probability $P(t \mid x_0)$, which denote the occupancy probability that an initially free particle that started from a point $x_0$ is bound to the target at time $t$; the occupancy probability $P(t \mid x_0)$ includes an additional step of the first-passage to the target. The expression $\tilde{P}(p \mid x_0)$ is derived as the Laplace transform of $P(t \mid x_0)$:

$$\tilde{P}(p|x_0) = \frac{\tilde{H}(p|x_0)}{p + k_{off}(1 - \tilde{H}(p))}, \tag{47}$$

where $\tilde{H}(p|x_0)$ is the Laplace transform of the probability density $H(t|x_0)$, which describes the first-passage time to the target when the particle started from point $x_0$. Since the achievement level is $K \geq 0$, the characteristic function ($CF$) is replaced by the Laplace transform in the form:

$$L(p) = \int_0^\infty e^{-pt} \frac{dP(t \mid x_0)}{dt} dt ,$$

which is equal to $L(p)=\int_0^\infty e^{-pt}\frac{dP(t\mid x_0)}{dt}dt = p\tilde{H}(p|x_0)-P(t=0|x_0)$; $P(t=0|x_0)$=0. Then:

$$L(p,x_0)=\frac{p\tilde{H}(p|x_0)}{p+k_{off}(1-\tilde{H}(p)}. \tag{48}$$

Recently, Lawley and Madrid proposed an elegant approximation in which the first binding time and the rebinding time $\tau$ after each unbinding event were assumed to follow an exponential distribution. The process *N(t)* could thus be approximated by a Markovian birth-death process, for which the distribution of the first-crossing time is known explicitly [86] (see also [87]). The first-binding time $\tau_0$ and the subsequent rebinding times $\tau_1$, $\tau_2$, ... of any particle are random variables, characterized by the survival probabilities $S(t\mid x_0)=P_{x_0}\{\tau_0>t\}$ and $S(t)=P\{\tau_i>t\}$, where $x_0$ is the starting point of the particle, and $P\{...\}$ denotes the probability of the random event within the braces.

Lawley and Madrid proposed an approximation, which relied on the approximation of these probabilities by an exponential function: $S(t\mid x_0)\approx S(t)\approx e^{-vt}$, with an appropriate rate $v$ [86]. In work [30], the relation is written, $\tilde{P}(p)=\frac{1}{|\Gamma|}\int_\Gamma dx_0\tilde{P}(p|x_0)$, where $|\Gamma|$ is the Lebesgue measure of the target region $\Gamma$ (e.g., the area of $\Gamma$ in the three-dimensional case). If the starting point $x_0$ is uniformly distributed, $P(t\mid x_0)$ should be replaced by $P(t\mid\circ)\equiv\frac{1}{|\Omega|}\int_\Omega dx_0 P(t|x_0)$, where $\circ$ indicates the uniform starting point. $\tilde{P}(p\mid\circ) = \tilde{H}(p\mid\circ)/(p+k_{off}(1-\tilde{H}(p))$, (by uniformly distribution for $x_0$) for $\tilde{H}(p)$ use Lawley-Madrid approximation (*LMA*), Ref. [86], where $H_{K,N}(t)$ is the probability density of the $T_{K,N}$, $\tilde{H}(p)$ is the Laplace transform of the probability density of the rebinding time τ, $H(t)=\frac{1}{|\Gamma|}\int_\Gamma dx_0 H(t|x_0)$;

$$L(p)=\frac{p\tilde{H}(p)}{p+k_{off}(1-\tilde{H}(p)}. \tag{49}$$

After binding, each particle remains in the target region $\Gamma$ for a random, exponentially distributed waiting time, characterized by the unbinding rate $k_{off}$, and then resumes diffusion from a uniformly distributed point on $\Gamma$. Expressions (47)-(48) are rewritten as:

$$\tilde{P}(p)=\frac{\tilde{H}(p)}{p+k_{off}(1-\tilde{H}(p)}, \qquad L(p)=\frac{p\tilde{H}(p)}{p+k_{off}(1-\tilde{H}(p)}. \tag{50}$$

To find the explicit form of the Laplace transform *L(p)* in Eqs. (49)-(50), we need to know the function $\tilde{H}(p)$. In work [30], consider for a shell-like domain $\Omega=\{x\in R^3:\ \rho<|x|<R\}$ bounded between two concentric spheres of radii $\rho$ and *R*. The inner sphere is a partially reactive target with reactivity $\kappa$, whereas the outer sphere is reflecting; then:

$$\tilde{H}(p|x_0)=\frac{g(r)}{g(\rho)-g'(\rho)D/\kappa}, \tag{51}$$

where *D* is the diffusion coefficient,

$g(r) = \dfrac{R\sqrt{p/D}\cosh\xi - \sinh\xi}{r\sqrt{p/D}}$, $g'(r) = \dfrac{(1 - Rrp/D)\sinh\xi - \xi\cosh\xi}{r^2\sqrt{p/D}}$, $\xi = (R - r)\sqrt{p/D}$,

$\kappa = k_{on}/(|\Gamma| N_A)$, $k_{on}$ is of the forward (bimolecular) reaction rate, $|\Gamma|$ is the surface area of the target and $N_A$ is the Avogadro number. Substituting this expression into Eq. (50) leads to a complex transcendental equation for the roots of the Lundberg equation (8).

Let's consider *LMA* valid for long-time, when $\tilde{H}(p) = \dfrac{\nu}{\nu + p}$. Equate $\ln L(p)$ to *k(p)* and substituting in Eq. (8). For *k(p)=s*, we get only one negative and one zero root.

In Ref. [30] consider short-time asymptotic behavior of Eq. (51), when:

$$\tilde{H}(p \mid \rho) \approx (\beta\sqrt{s_1} - 1)/[(\nu + \beta)\sqrt{s_1} - (1 + \nu) + \gamma s_1], \tag{52}$$

$$s_1 = (R-\rho)^2 p/D, \quad \gamma = \nu\beta^2\rho/R, \quad \beta = (R-\rho)/\rho,$$

$$\tilde{H}(p|r) \approx (\rho/r)\, e^{-\sqrt{s_1}(r-\rho)/(R-\rho)}\tilde{H}(p \mid \rho). \tag{53}$$

Let's consider the case when $(r-\rho)/(R-\rho) << 1$, $e^{-\sqrt{s_1}(r-\rho)/(R-\rho)} \approx 1 - \sqrt{s_1}(r-\rho)/(R-\rho)$.

Substituting in this approximation expressions (52)-(53) into Eqs. (49)-(50) and taking into account that equation (8) takes the form $\ln L(p) = k(p) = s$, we obtain a positive solution to this equation in the form:

$$\rho_+(s) = \frac{K_1^2}{K_2^2}[1 + \frac{2K_2(1+\nu-\rho e^{-s}/r)}{K_1^2} - \sqrt{1 + \frac{4K_2(1+\nu-\rho e^{-s}/r)}{K_1^2}}], \tag{54}$$

$$K_1 = e^{-s}\frac{\rho}{r}(\sqrt{\beta} + \frac{r-\rho}{R-\rho}) - (\nu+\beta), \quad K_2 = e^{-s}\frac{\rho}{r}\beta\frac{r-\rho}{R-\rho} + \gamma.$$

Substituting Eq. (54) into Eqs. (43)-(45) and setting *x*=*K*=2, we obtain an expression that shows the average time the system stays above the level *K*=2. Below this level, reactions no longer occur, since they require at least two particles. Here we consider *N* particles that independently diffuse with diffusion coefficient *D* inside a bounded domain $\Omega \in R_d$ with a smooth boundary $\partial\Omega$ that is reflecting everywhere except for a target region *Γ* with a finite reactivity *κ*, *ν* is appropriate rate [85]. In order to get the correct long-time behavior of the survival probability, one can set $\nu = D \cdot \lambda_1$ to match the leading term of the exact spectral expansion. Alternatively, as the rebinding time *τ* is approximated by an exponential law, one can set $\nu = 1/\langle\tau\rangle$. When the target is small and weakly reactive, equation $\lambda_1 \approx \kappa|\Gamma|/D|\Omega| = 1/D\langle\tau\rangle$ indicates that $\nu = 1/\langle\tau\rangle$ is close to $D \cdot \lambda_1$, and both choices yield the same result. In fact, as restricted diffusion occurs in a bounded domain, the governing Laplace operator, −Δ, has a discrete spectrum, i.e., a countable set of eigenvalues $0 < \lambda_1 \le \lambda_2 \le \lambda_3 \le \ldots \nearrow \infty$ associated to eigenfunctions $\{v_n^{(p)}(\mathbf{x})\}$ forming an orthonormal basis of the space $L_2(\partial\Omega)$ of square integrable functions on $\partial\Omega$.

But expressions (43)-(45) are valid for *k(s)* of the form Eq. (26) for case Eq. (12). Probably, in some approximation (which must be estimated) they are also valid for *k(s)* of the form Eqs. (49)-(51). A more general expression for the distribution of the time of stay at level *x MGF* $D_x^+(\mu) = \boldsymbol{E}e^{-\mu Q_x(\infty)}$ has a more complex form. In general, the transformation is defined [12], for example, $\int_{-0}^{+\infty} e^{i\alpha x} dx D^{\pm}_x(s,\mu)$, where $D^{\pm}_x(s,\mu) = D_x(s,\mu) I(\pm x > 0)$. The inverse transformation, for example, for the process $\xi(t) = at + w(t)$, where $w(t)$ is a Brownian motion process, has the form:

$$D^{+}{}_{x}(s,\mu) = \boldsymbol{E}e^{-\mu Q_x(\theta_s)} = 1 - \frac{\rho_+(s+\mu) - \rho_+(s)}{\rho_+(s+\mu)} e^{-\rho_+(s)x}, \quad x > 0, \tag{55}$$

$$D^{-}{}_{x}(s,\mu) = \frac{s}{s+\mu} - \frac{\rho_+(s)}{2(s+\mu)} e^{-\rho_-(s+\mu)x} (\rho_-(s) - \rho_-(s+\mu)), \quad x < 0.$$

Similar relationships are also written for a similar diffusion-controlled process [30] (or death and birth in Lawley and Madrid an approximation (*LMA*)). The average value of $Q_x(\theta_s)$ is:

$$\boldsymbol{E}Q_x(\theta_s) = -\frac{\partial D^{+}{}_{x}(s,\mu)}{\partial \mu}\bigg|_{\eta=0} = \frac{\partial \rho_+(s)}{\partial s} \frac{1}{\rho_+(s)} e^{-\rho_+(s)x}, \tag{56}$$

where $\rho_+(s)$ is defined in Eq. (54).

More general expressions for other boundary functionals, for example, for extrema, are also written down. The moment of the first achievement of the maximum of the process $\{\xi(t), \xi(0)=0, t \geq 0\}$ on the interval $[0,t]$, equal to $T_+(t) := \{u > 0 : \xi(u) = \xi^+(t)\}$ Eq. (0.7), was not considered above. The relation [12] is valid:

$$\boldsymbol{E}[e^{-\mu T_+(\theta_s)} / \xi^+(\theta_s) > 0] = \frac{q_+(s+\mu)}{q_+(s)} D_{+0}(s,\mu), \;\; D_{+0}(s,\mu) = \frac{p_+(s)}{p_+(s+\mu)}, \quad p_+(s) = P\{\xi^+(\theta_s) = 0\}. \tag{57}$$

Equation (8) for this case has a single positive root $\rho_+(s)$ for $s$>0, $q_+(s) = 1 - p_+(s)$. Substituting Eq. (54) into Eqs. (55)-(57) leads to explicit expressions for the desired quantities.

### 5.3. Consideration of the unicyclic network with affinity *A* in periodically controlled systems without using approximations

For the process of the unicyclic network with affinity *A* discussed in Section 4.1, we consider the *SCGF* of the form Eq. (25) without employing the approximations that lead to relation Eq. (26). The Lundberg equation (8) in this case takes the form:

$$k^+\left[e^{r/N} + e^{-(r+A)/N} - 1 - e^{-A/N}\right] = s. \tag{58}$$

Solving equation (58), we obtain the expressions:

$$\rho_+(s) = N\ln[\frac{1}{2}(1+e^{-A/N}+\frac{s}{k^+}) + \sqrt{\frac{1}{4}(1+e^{-A/N}+\frac{s}{k^+})^2 - e^{-A/N}}], \tag{59}$$

$$\rho_-(s) = -N\ln[\frac{1}{2}(1+e^{-A/N}+\frac{s}{k^+}) - \sqrt{\frac{1}{4}(1+e^{-A/N}+\frac{s}{k^+})^2 - e^{-A/N}}].$$

Substituting into Eq. (59) the numerical values of the parameters used in Section 4.1, *N*=5, *A*=-1, $k^-$=1, *λ*=0.819, *J*=0.36, *c*=0.2, and *m*=-0.036, we obtain:

$$\rho_+(s) = 5\ln[\frac{1}{2}(2.2214+\frac{s}{0.819}) + \sqrt{\frac{1}{4}(2.2214+\frac{s}{0.819})^2 - 1.2214}], \tag{60}$$

$$\rho_-(s) = -5\ln[\frac{1}{2}(2.2214+\frac{s}{0.819}) - \sqrt{\frac{1}{4}(2.2214+\frac{s}{0.819})^2 - 1.2214}].$$

Fig. 8 shows the behavior of the quantities Eq. (60). A significant difference from Fig. 1 is obvious.

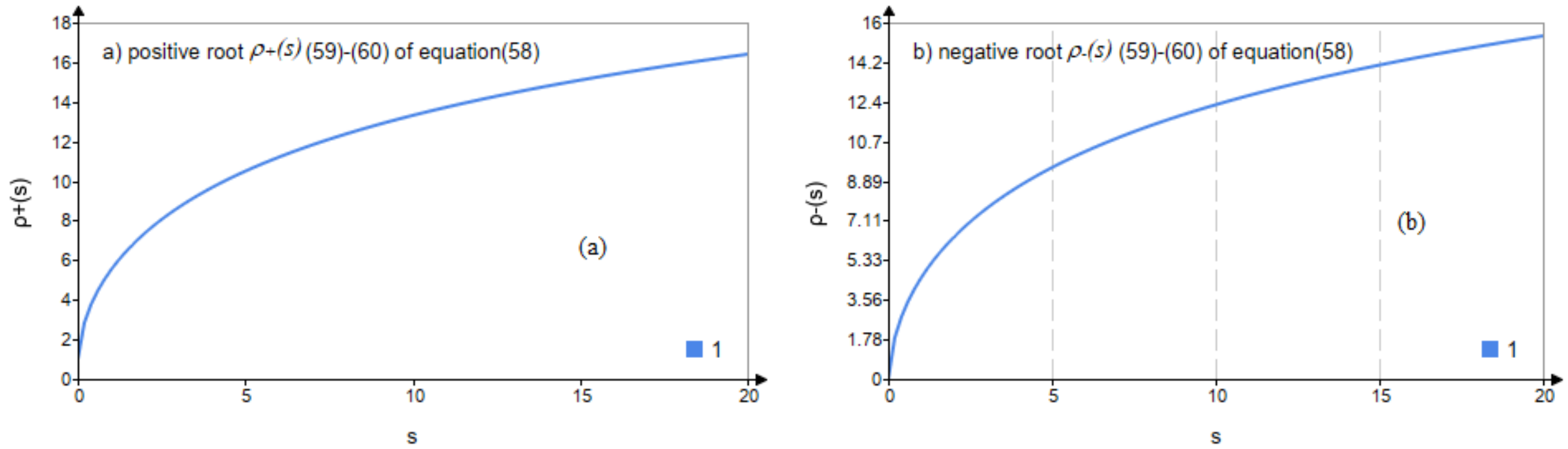

Fig.8. Behavior of functions $\rho_+(s)$ Fig.8a), $\rho_-(s)$ Fig.8b) of type Eq. (60).

But, despite the difference in the values $\rho_+(s)$, $\rho_-(s)$, the behavior of the dependencies of the average *FPT* on *s* differs slightly only numerically, maintaining the characteristic behavior (Fig. 2b, Fig. 9a).

In the works [12], [13] an expression for the *MGF* $T(s,x)=\boldsymbol{E}[e^{-s\tau^+(x)},\tau^+(x)<\infty]$ of the form was obtained:

$$T(s,x)=\boldsymbol{E}[e^{-s\tau^+(x)},\tau^+(x)<\infty]=e^{-\rho_+(s)x},\quad x>0, \tag{61}$$

$$\boldsymbol{E}[\tau^+(x),s]=-\frac{\partial \ln \boldsymbol{E}[e^{-s\tau^+(x)},\,\tau^+(x)<\infty]}{\partial s}=\frac{\boldsymbol{E}[\tau^+(x)e^{-s\tau^+(x)},\,\tau^+(x)<\infty]}{\boldsymbol{E}[e^{-s\tau^+(x)},\,\tau^+(x)<\infty]}=x\frac{\partial\rho_+(s)}{\partial s}.$$

Figure 9a shows the results of comparison of calculations carried out using expressions (29), (30) and Eq. (61), as well as using *CSGF* Eq. (25) and the approximation Eq. (26) obtained from it at *x*=*u*=0.08.

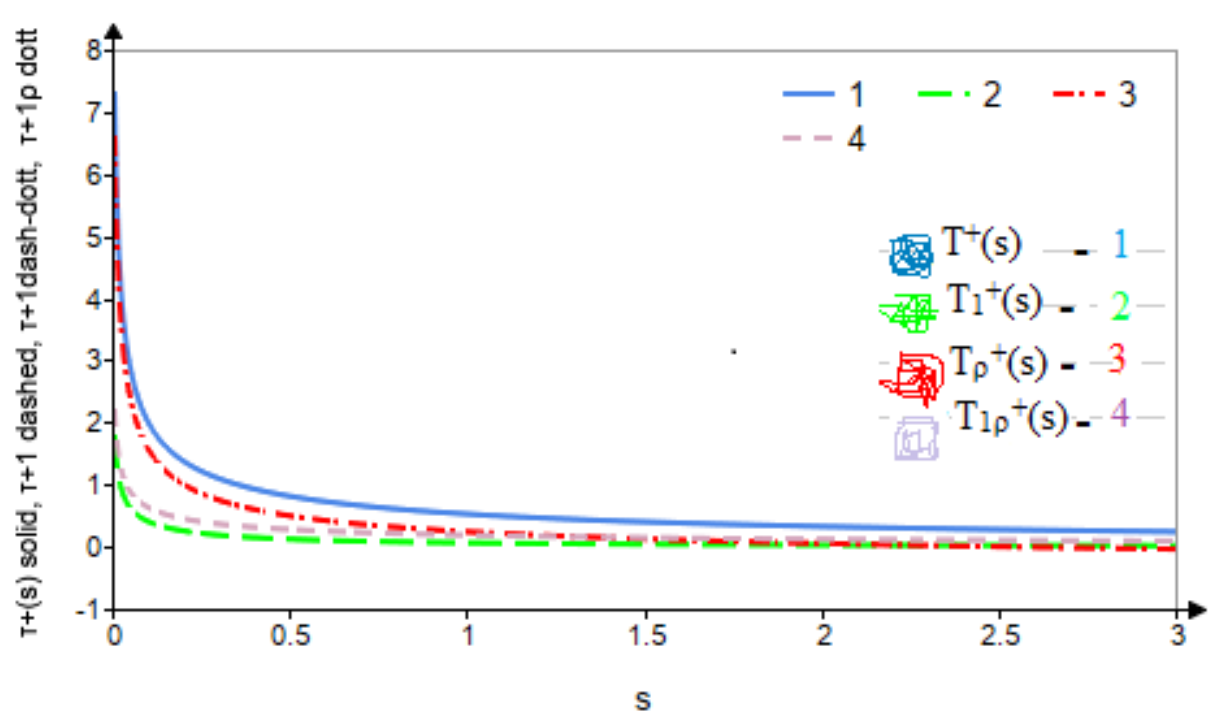

Fig.9a. Comparison of the dependencies of the average *FPT* on *s* at *x*=*u*=0.08. 1). $T^+(s)$, solid, blue, Fig. 2b, obtained from Eqs. (12), (26)-(28) and (30); 2). $T_1^+(s)$, dashed, green, obtained from Eqs. (26)-(28) and (61); 3). $T_\rho^+(s)$, dash-dotted, red, obtained from Eqs. (30) and (59)-(60); 4). $T_{1\rho}^+(s)$, dotted, pink, obtained from Eqs. (59)-(61).

From Fig.9a it is evident that the dependencies at *x*=*u*=0.08 $T^+(s)$, $T_1^+(s)$, $T_\rho^+(s)$, and $T_{1\rho}^+(s)$ are close to each other.

However, the results of Fig. 9a provide only a very rough approximation. In Fig. 9a, expression (17) is used, which is valid for the exponential distribution of fluctuations $\overline{F}(x)$ of the form Eq. (12). For expression (25), the distribution function of the magnitude of the jumps $\overline{F}(x)$ has the form:

$$\bar{F}(x)=\begin{cases}1, & x<-1/N\\ \delta(x)(c/k^{+})/(1+c/k^{+}\delta(x)), & -1/N<x<0\\ e^{-A/N}/(1+c/k^{+}\delta(x)), & 0\le x\le 1/N\\ 0, & x>1/N\end{cases}.$$

Using expressions:

$$q_{+}(s,u)=\boldsymbol{E}[e^{-s\tau^{+}(u)},\tau^{+}(u)<\infty]=\int_{-0}^{u}G_{0}(s,u-y)dP_{+}(s,y),$$

$$G_{0}(s,u)=\frac{\lambda\rho_{-}(s)}{s}\int_{0}^{\infty}e^{-\rho_{-}(s)x}\bar{F}(u+x)dx=\frac{\lambda\rho_{-}(s)}{s}[(1+k^{+}e^{-A/N}-k^{+}-k^{+}e^{-A/N+\rho_{-}(s)(u-1/N)})/\rho_{-}(s)-c],$$

$$q_{+}(s,u)=\frac{\lambda}{c}\int_{0}^{\infty}e^{-\rho_{-}(s)x}\bar{F}(u+x)dx+\frac{\lambda\rho_{-}(s)}{s}\int_{0+}^{u}\int_{0}^{\infty}e^{-\rho_{-}(s)y}\bar{F}(u-y+x)dxdP_{+}(s,y),$$

$$dP_{+}(s,y)=-dq_{+}(s,u)\quad(y>0),$$

written in [12] and valid for arbitrary functions $\bar{F}(x)$ of process (2), we obtain for $MGF$ $q_{+}(s,u)$ in this case the expressions:

$$q_{+}(s,u)=\frac{(0.819\cdot 1.2214)[0.819\cdot 1.2214+se^{-f(s)u}][1-e^{-0.2\rho_{-}(s)}]}{0.2[s+0.819\cdot 1.2214]\rho_{-}(s)},\quad u\le 1/N,$$

$$f(s)=-\rho_{-}(s)\frac{(s+\lambda e^{-A/N})}{s+\lambda e^{-A/N}(1-e^{-\rho_{-}(s)/N})},$$

$$\boldsymbol{E}[\tau^{+}(u),s]=\boldsymbol{E}[\tau^{+}(u)e^{-s\tau^{+}(u)},\tau^{+}(u)<\infty]/\boldsymbol{E}[e^{-s\tau^{+}(u)},\tau^{+}(u)<\infty]=-\partial\ln q_{+}(s,u)/\partial s=$$

$$=\frac{-e^{-f(s)u}(1-su\partial f(s)/\partial s))}{0.819e^{-A/N}+se^{-f(s)u}}-\partial\rho_{-}(s)/\partial s\frac{0.2e^{-0.2\rho_{-}(s)}}{1-e^{-0.2\rho_{-}(s)}}+\frac{\partial\rho_{-}(s)/\partial s}{\rho_{-}(s)}+\frac{1}{s+0.819e^{-A/N}}.$$

The results of calculating the average value of the function $\tau^{+}$ using this expression with a function $\rho_{-}(s)$ of the form (60) are shown by a solid line in Figs. 9b, c. They differ from the results of calculation using expression (30), Fig. 9a. Expression (17) is obtained from these expressions by specifying distribution (12) and using the relations $c\rho_{-}(s)p_{+}(s)=s,\quad \rho_{+}(s)=bp_{+}(s)$ valid for case Eq. (12). The dependencies constructed using the obtained expression differ from Fig. 9a and have the form:

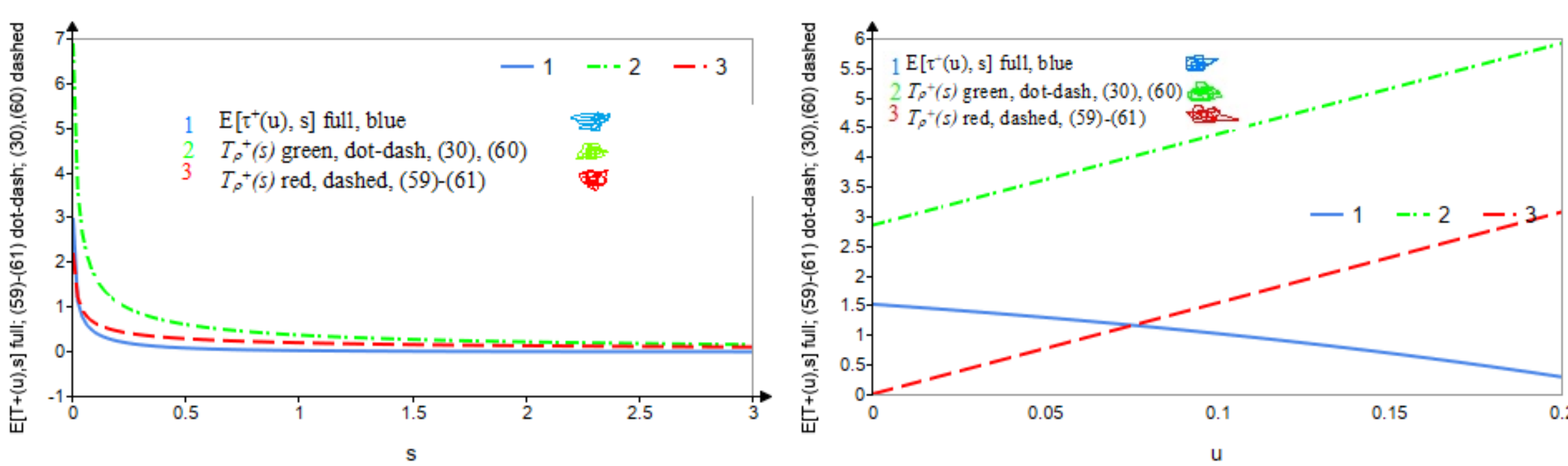

Fig. 9 b. Dependence curves $\boldsymbol{E}[\tau^{+}(u),s]$ on s for a fixed value of $u$=0.08, calculated using the above-mentioned relationships characterizing expression (25) (solid blue line), using expressions (30) and (60) (green, dot-dashed), in the interval $s$=0,…,3, and calculated using expression (61) (red dotted line) for functions Eqs. (59)-(60), Fig. 8.

Fig. 9c. Curves of dependencies $\boldsymbol{E}[\tau^{+}(u),s]$ on $u$ in the interval $u$=0,…,0.2 at a fixed value $s$=0.02, calculated according to the above relations characterizing expression (25) (solid blue line), using expressions (30) and (60) (green, dot-dashed), and calculated according to expression (61) (red dotted line) for functions Eqs. (59)-(60), Fig. 8.

The value of $u$ was chosen fixed at $u$=0.08. The dependence on $u$ at a fixed value of $s$=0.02 is shown in Fig. 9c. It is evident that the curves $\boldsymbol{E}[\tau^{+}(u),s]$ differ. The interval of change of the value $u$ can be increased by considering the value $m/N$ instead of $1/N$, $m$=1,2,3…. The argument $r$ in expression (25) is replaced by $mr$. We do not dwell on the details of obtaining calculation formulas and the influence of changing parameters on the calculation results.

### 5.4. Two-level model. Accounting for dependence on entropy change

Let us consider an example of a process unrelated to the claim surplus process $\zeta(t)$, Eq. (2). In paper [61], the influence of entropy changes on the *FPT* process in thermodynamics of trajectories is considered. One example chosen is a two-level model. In works [79, 91], a classical two-level system is described. The classical stochastic system is described by the Master Equation $\partial_t|P(t)\rangle = W|P(t)\rangle$ [92]. The vector $|P(t)\rangle$ represents the probability distribution at time $t$, $|P(t)\rangle = \sum_C P(C,t)|C\rangle$, where $P(C,t)$ describes the probability that the system is in configuration $C$ at time $t$, and $\{|C\rangle\}$ is the orthonormal basis of the configuration, $\langle C|C'\rangle = \delta_{CC'}$. The master operator $W$ is the matrix $W = \sum_{C'\neq C} w(C\to C')|C'\rangle\langle C| - \sum_C R(C)|C\rangle\langle C|$. Here $w(C\to C')$ is the rate of transition from $C$ to $C'$, and $R(C) = \sum_{C'} w(C\to C')$ is the rate of exit from $C$. In this description, the expectation of the operator $T$ is defined as $\langle T(t)\rangle = \langle -|T|P(t)\rangle$, where $\langle -| \equiv \sum_C \langle C|$ (such that $\langle -|P(t)\rangle = 1$ due to conservation of probability). The operator $W_s$ is given by $W_s = \begin{pmatrix} -\eta & e^{-s}\kappa \\ e^{-s}\eta & \kappa \end{pmatrix}$, where $W_s = \sum_{C'\neq C} e^{-s} w(C\to C')|C'\rangle\langle C| - \sum_C R(C)|C\rangle\langle C|$ [74-79, 92], $W_s$ is a deformed operator $W$, and $\theta(s)$ Eq. (62) is its largest eigenvalue; $\eta$ and $\kappa$ are transition rates. In a two-level system with only two configurations, $C\in\{0,1\}$, and the transition rates are $w(0\to 1)=\kappa$ and $w(1\to 0)=\eta$. From $\theta(s)$, the activity $K$ can be calculated: $\boldsymbol{E}K/\tau = -\partial\theta(s)/\partial s_{|s=0}$.

The operator $T_x$ is a transfer matrix operator $T_x = \sum_{C`\neq C} \frac{w(C\to C`)}{x+R(C)}|C`\rangle\langle C|$. For this problem is given by $T_x = \begin{pmatrix} 0 & \kappa/(x+\kappa) \\ \eta/(x+\eta) & 0 \end{pmatrix}$, and from its largest eigenvalue, we obtain the large deviation function $g(\gamma)$, Eq. (62). The functions $g(\gamma)$ and $\theta(s)$ are introduced before Fig.1. This function $g(\gamma)$ is the inverse of the function $\theta(s)$, Eq. (62).

For a classical two-level system [79, 91] as described in work [91], for *SCGF* $g(\gamma)$ and $\theta(s)$ the following expressions are obtained:

$$g(\gamma) = -\ln(1+\gamma a_1 + \gamma^2 a_2),\ \ a_1 = \frac{\eta+\kappa}{\eta\kappa},\ \ a_2 = \frac{1}{\eta\kappa};$$
$$\theta(s) = \frac{1}{2}[\sqrt{(\eta-\kappa)^2 + 4\eta\kappa e^{-s}} - (\eta+\kappa)]. \tag{62}$$

Function $\theta(s=-r)=k(r)$ (*k(r)* from Eqs. (7)-(8)), this function appears in the Lundberg equation (8). The solution of this equation (of the form $\theta(s=-r)=k(r)=s$ ) has the form $r=\ln(1+sa_1+s^2a_2)=-g(s)>0$. In this case $r=r_+=\rho_+(s)=-g(s)$, as noted before Fig.1 (that $-g(s)=\rho_+(s)$). *FPT* are considered in paper [61]. Such a consideration was carried out in Section 4.

Let us now consider a two-level model using lattice Poisson processes [12]. A homogeneous process with independent increments $\{\xi(t), t\geq 0, \xi(0)=0\}$ is called a lattice Poisson process if in the characteristic function in a Lévy-Khinchin type representation Eq. (0.4) *a*=0 and the Lévy measure of jumps is concentrated in a sequence of points $\{x_r=rh\}$ ( $h>0, r\neq 0$ are integer values) and is defined by the relation:

$$\Pi(x_r)=\lambda p_r,\quad p_r=P\{\xi_1=rh\},\quad r=\pm 1,\pm 2,\ldots .$$

The *MGF* of such a process is determined by the relation:

$$\tilde{q}_t\{z\}:=\boldsymbol{E}z^{\xi(t)}=e^{tk(z)},\quad |z|=1,$$

$$k(z)=\lambda(p(z)-1),\quad p(z)=\boldsymbol{E}z^{\xi_1}=\sum_{r\neq 0}z^{rh}p_r .$$

Without loss of generality, we can assume *h*=1 and consider only integer processes. We call the function $k(z)=\ln\tilde{q}_1\{z\}$ the cumulant (or *SCGF*) of the process. If for an integer Poisson process $\xi(t)=\sum_{k\leq\nu(t)}\xi_k$, $\xi_0=\xi(0)=0$, where $\nu(t)$ is a simple Poisson process with parameter $\lambda>0$, the conditional *MGF* of jumps satisfies the condition:

$$\boldsymbol{E}[z^{\xi_1}/\xi_1<0]=z,\ |z|\leq 1,$$

then the process $\xi(t)$ is called upper semicontinuous.

By making the substitution $s=\ln z$ in expression (62), we obtain *SCGF* of the form:

$$\theta(z)=\frac{1}{2}[\sqrt{(\eta-\kappa)^2+4\eta\kappa z^{-1}}-(\eta+\kappa)],\quad k(z)=\frac{1}{2}[\sqrt{(\eta-\kappa)^2+4\eta\kappa z}-(\eta+\kappa)]. \tag{63}$$

The positive root $z_s>1$ of the Lundberg equation $k(z)=s$ is:

$$z_s=1+sa_1+s^2a_2=e^{-g(s)}, \tag{64}$$

where the function $g(s)$ is written in Eq. (62). For functionals associated with the intersection of an integer level, an integer process $\xi(t)$ can hit a fixed integer level with positive probability. Therefore, in work [12] notations of this type are used (*x*>0):

$$\tau^+(x)=\inf\{t>0:\xi(t)>x\},\quad \overline{\tau}^+(x)=\inf\{t>0:\xi(t)\geq x\}.$$

In work [12] theorem 7.4 is proved, according to which for an upper semicontinuous process $\xi(t)$ the *MGF* $\tau^+(x)$ ( $\overline{\tau}^+(x)$ ) are defined by the relations:

$$T(s,x)=\boldsymbol{E}e^{-s\tau^+(x)},\tau^+(x)<\infty]=P\{\xi^+(\theta_s)>x\}=q_+(s)^{x+1}, \tag{65}$$

$$\overline{T}(s,x)=\boldsymbol{E}[e^{-s\overline{\tau}^+(x)},\overline{\tau}^+(x)<\infty]=P\{\xi^+(\theta_s)\geq x\}=q_+(s)^x,$$

$$p_k^+=\overline{T}(s,k)-T(s,k)=p_+(s)q_+(s)^k,\quad q_+(s)=z_s^{-1},$$

where $z_s>1$ is the positive root of the Lundberg equation $k(z)=s$, and the *MGF* of $\xi^+(\theta_s)$ determined by the relation:

$$\tilde{q}_+(s,z)=\boldsymbol{E}z^{\xi^+(\theta_s)}=\frac{p_+(s)}{1-q_+(s)z}. \tag{66}$$

From expressions (64)-(65) we find that:

$$q_+(s) = \frac{1}{1+sa_1+s^2a_2}, \qquad p_+(s) = 1 - q_+(s) = \frac{sa_1+s^2a_2}{1+sa_1+s^2a_2}. \tag{67}$$

Substituting relation (67) into expression (65), we find that:

$$\boldsymbol{E}\tau^+(x,s) = -\frac{\partial \ln T(s,x)}{\partial s} = \frac{\boldsymbol{E}[\tau^+(x)e^{-s\tau^+(x)}, \tau^+(x)<\infty]}{\boldsymbol{E}[e^{-s\tau^+(x)}, \tau^+(x)<\infty]} = -\frac{x+1}{q_+(s)}\frac{\partial q_+(s)}{\partial s} = (x+1)\frac{a_1+2sa_2}{1+sa_1+s^2a_2}. \tag{68}$$

In the article [61], the expression for this quantity is obtained using the relations of thermodynamics of trajectories from *SCGF g(γ)*. In the article [61] this relation is written as:

$$\langle \tau_\gamma \rangle \simeq -\langle K_0 \rangle \frac{\partial g(\gamma)}{\partial \gamma} = \langle K_0 \rangle \frac{a_1+2\gamma a_2}{1+\gamma a_1+\gamma^2 a_2}. \tag{69}$$

This ratio coincides with expression (68) when:

$$\boldsymbol{E}\tau^+(x,s) = \langle \tau_\gamma \rangle,\ \langle K_0 \rangle = \boldsymbol{E}K(s=0) = x+1,\ \ s=\gamma.$$

In the article [61], the parameter $s=\gamma$ is expressed through the total entropy change $\Delta s_{tot}$. The dependences of the mean value $\boldsymbol{E}\tau^+(x,s) = \langle \tau_\gamma \rangle$ and other moments of random variables on $\Delta s_{tot}$ are obtained.

From expression (66) we obtain:

$$\boldsymbol{E}\xi^+(\theta_s,z) = z\frac{\partial \ln \tilde{q}_+(s,z)}{\partial z} = z\frac{\boldsymbol{E}\xi^+(\theta_s)z^{\xi^+(\theta_s)-1}}{\boldsymbol{E}z^{\xi^+(\theta_s)}} = \frac{z}{1+sa_1+s^2a_2-z}. \tag{70}$$

Analogous expressions have been derived for the continuous case in Section 4.3. In order to express the quantity $\boldsymbol{E}\xi^+(\theta_s,z)$ in terms of the total change in entropy $\Delta s_{tot}$, as done for quantity Eq. (69) in reference [61], it is necessary to relate the parameter $z$ to $\Delta s_{tot}$. This requires incorporating the quantity $\xi^+(\theta_s)$ as a thermodynamic parameter into the distribution that characterizes the statistical properties of the system, thereby obtaining the entropy of this distribution, where the thermodynamic parameter conjugate to the quantity $\xi^+(\theta_s)$ will be $s_1 = \ln z$. The parameter $s_1$ can be expressed in terms of $\Delta s_{tot} = \Delta s_{sys} + \Delta s_m$, where $\Delta s_{sys}$ is the system's entropy change and $\Delta s_m$ is the entropy exchange with the environment.

The following relations are also valid [12]:

$$p_r(s) = P\{\xi(\theta_s)=r\} = sz_s^{-r-1}z_s' \ \ (k(z_s)=s,\ s>0,\ r>0),$$

$$P\{\xi(t)=r\} = \frac{t}{r}\frac{\partial}{\partial t}P\{\tau^+(r)<t/\tau^+(0)<t\} \ \ (r>0),$$

$$P\{\xi(t)=r\} = -\frac{t}{r}\frac{\partial}{\partial t}P\{\tau^-(r)<t/\tau^-(0)<t\} \ \ (r<0),$$

$$P\{\xi^+(\theta_s)\ge r\} = z_s^{-r}.$$

Relationships are also written for $\tilde{q}_-(s,z)$, for functionals $\gamma^+(x)$, Eq. (5), for the time spent in a certain fixed state, which are determined by the distribution of the process $\xi(\theta_s)$, for the time spent above level $x$, for the bankruptcy function and other relations.

### 5.5. Brownian motion and diffusion

Diffusion processes are closely related to the process of Brownian motion $\xi_0(t) = at + \sigma w(t)$, where $w(t)$ is a Wiener process. The process of Brownian motion is mathematical model of the diffusion process. The process of Brownian motion is continuous both from above and from below with *m*=*a*, $D\xi(1) = \sigma$. For the characteristic function $\xi(\theta_s)$, the following relation is valid (see expressions (7), (13)):

$$\varphi(s,\alpha) := \boldsymbol{E}e^{i\alpha\xi(\theta_s)} = \frac{s}{s-\Psi(\alpha)},$$

and also, the fundamental factorization identity (the identity of infinitely divisible factorization):

$$\varphi(s,\alpha) = \varphi_+(s,\alpha)\varphi_-(s,\alpha), \quad \operatorname{Im}\alpha = 0.$$

For the process of Brownian motion:

$$\varphi(s,\alpha) = \frac{2s}{2s - 2i\alpha a - (i\alpha)^2\sigma^2}, \quad \varphi_+(s,\alpha) = \frac{\rho_+(s)}{\rho_+(s) - i\alpha}, \quad \varphi_-(s,\alpha) = \frac{\rho_-(s)}{\rho_-(s) + i\alpha}, \tag{71}$$

$$\rho_\pm(s) = \frac{\sqrt{2s\sigma^2 + a^2} \mp a}{\sigma^2}, \quad \Psi(\alpha) = i\alpha a - \sigma^2\alpha^2/2, \quad \Psi(\alpha)_{|i\alpha=r} := k(r) = ar + \sigma^2 r^2/2 = s.$$

Fig.10 shows the behavior of functions $\rho_+(s)$ Fig.10a), $\rho_-(s)$ Fig.10b) from expression (71) for $\sigma$=1, *a*=-0.036. It is evident that the behavior of functions $\rho_+(s)$ Fig.10a), $\rho_-(s)$ Fig.10b) from expression (71) is similar to the behavior of functions $\rho_+(s)$ Fig.8a), $\rho_-(s)$ Fig.8b) from Eq. (60) and differs from the behavior of functions $\rho_+(s)$ Fig.1a), $\rho_-(s)$ Fig.1b) from Eq. (28).

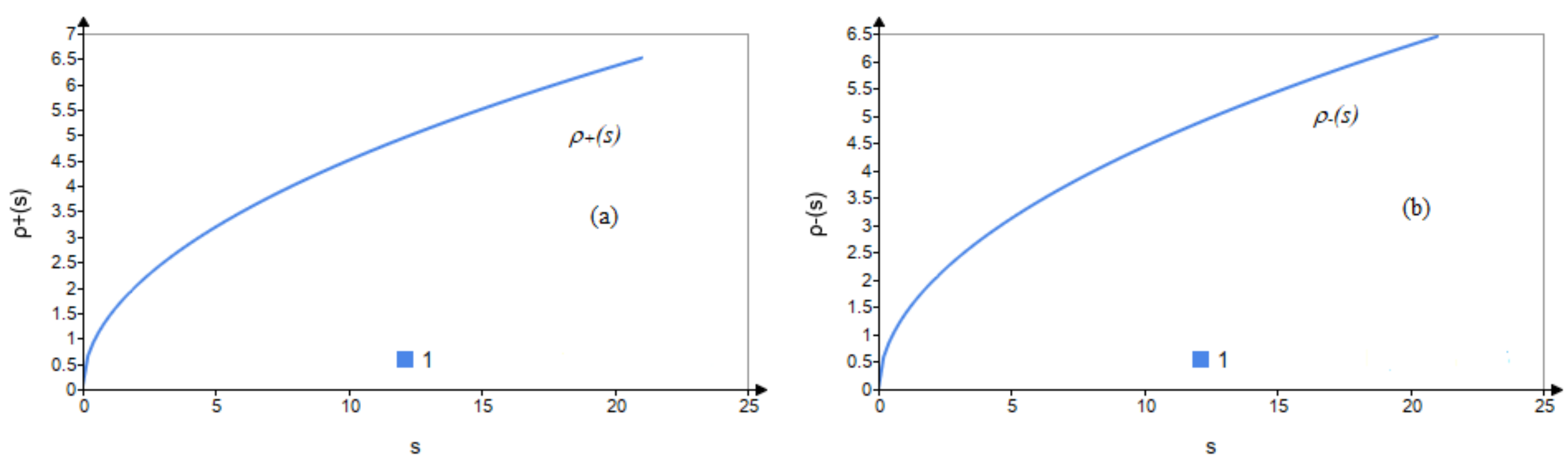

Fig.10. Behavior of functions $\rho_+(s)$ Fig.10a), $\rho_-(s)$ Fig.10b) of type Eq. (71).

The relation [12] is fulfilled:

$$\varphi_\pm(s,\alpha) := \boldsymbol{E}e^{i\alpha\xi^\pm(\theta_s)}. \tag{72}$$

From Eqs. (71)-(72) we obtain at $i\alpha = \mu$:

$$\boldsymbol{E}e^{\mu\xi^\pm(\theta_s)} = \frac{\sqrt{2s\sigma^2 + a^2} \mp a}{\sqrt{2s\sigma^2 + a^2} \mp a \mp \mu\sigma^2}, \tag{73}$$

$$\boldsymbol{E}[\xi^\pm(\theta_s)e^{\mu\xi^\pm(\theta_s)}] = \frac{\partial\varphi_\pm(s,\alpha=-i\eta)}{\partial\mu} = \sigma^2\frac{(\pm)\sqrt{2s\sigma^2 + a^2} \mp a}{(\sqrt{2s\sigma^2 + a^2} \mp a \mp \mu\sigma^2)^2}_{|\mu=0} = \frac{(\pm)\sigma^2}{\sqrt{2s\sigma^2 + a^2} \mp a},$$

$$\boldsymbol{E}\xi^\pm(\theta_s,\mu) = \frac{\boldsymbol{E}[\xi^\pm(\theta_s)e^{\mu\xi^\pm(\theta_s)}]}{\boldsymbol{E}e^{\mu\xi^\pm(\theta_s)}} = \frac{\partial\ln\varphi_\pm(s,\alpha=-i\eta)}{\partial\mu} = \frac{(\pm)\sigma^2}{\sqrt{2s\sigma^2 + a^2} \mp a \mp \mu\sigma^2}, \tag{74}$$

$$\boldsymbol{E}\xi^{+}(\theta_s,\mu)=\frac{\sigma^2}{\sqrt{2s\sigma^2+a^2}-a-\mu\sigma^2},\quad \boldsymbol{E}\xi^{-}(\theta_s,\mu)=\frac{-\sigma^2}{\sqrt{2s\sigma^2+a^2}+a+\mu\sigma^2}.$$

From the dependencies of a randomly stopped process on an exponentially distributed random variable $\theta_s$, Eq. (0.5), one can move to dependencies on time using the inverse Laplace-Carson transform. For example, the inverse Carson-Laplace transform of expression (74) has the form:

$$\boldsymbol{E}\xi^{+}(t,\mu)=\frac{1}{\mu\sigma^2(\mu\sigma^2+2a)}[-(\mu\sigma^2+a)e^{-t\mu(\mu\sigma^2+2a)/2}erfc(-(\mu\sigma^2+a)\sqrt{t/2\sigma^2})-a\cdot erf(a\sqrt{t/2\sigma^2})-(\mu\sigma^2+a)],$$

$\mu>-a/\sigma^2$.

Dependencies of type Eq. (74) on the parameter $\mu$ at a fixed value of $s$, and on the parameter $s$ at a fixed value $\mu$ are shown in Fig. 11.

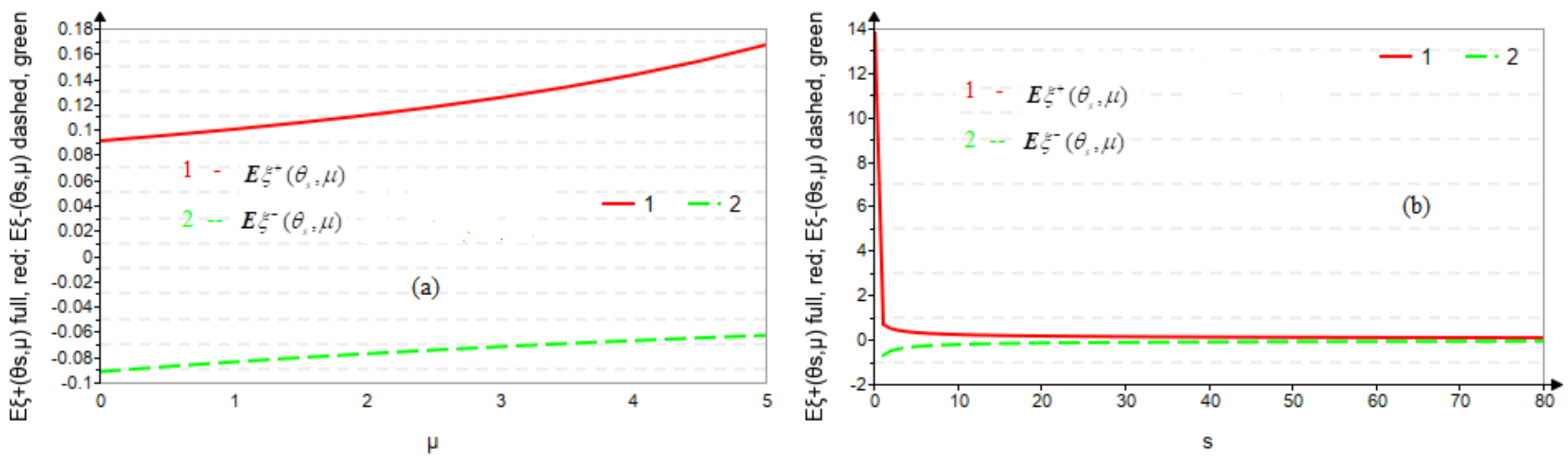

Fig.11. Fig.11a): Dependencies of type Eq. (74) of values $\boldsymbol{E}\xi^{+}(\theta_s,\mu)$ (full, red) and $\boldsymbol{E}\xi^{-}(\theta_s,\mu)$ (dashed, green) on the parameter $\mu$ at a fixed value of $s$=60; Fig.11b): Dependencies of type Eq. (74) of values $\boldsymbol{E}\xi^{+}(\theta_s,\mu)$ (full, red) and $\boldsymbol{E}\xi^{-}(\theta_s,\mu)$ (dashed, green) on the parameter $s$ at a fixed value $\mu$=0 at $\sigma$=*1, a=-0.036.*

To determine the average value of *FPT*, we use two relations: expression (61) and the relation written in [12], which includes the boundary functional of jumping over level $x$, of the form $\gamma^{+}(x)=\xi(\tau^{+}(x))-x$ (the first overjump over $x$>0, the severity of ruin), defined in expression (5). This *MGF* has the form:

$$\boldsymbol{E}[e^{-s\tau^{+}(x)-u\gamma^{+}(x)},\tau^{+}(x)<x]=\frac{cq_{+}(s)}{c+u}e^{-\rho_{+}(s)x},\ \ x>0,\ \ c>0,\ \ q_{+}(s)=1-\rho_{+}(s)/c,\qquad (75)$$

$$\boldsymbol{E}\tau^{+}(x,s)=\frac{\boldsymbol{E}[\tau^{+}(x)e^{-s\tau^{+}(x)},\tau^{+}(x)<x]}{\boldsymbol{E}[e^{-s\tau^{+}(x)},\tau^{+}(x)<x]}=-\frac{\partial\ln\boldsymbol{E}[e^{-s\tau^{+}(x)-u\gamma^{+}(x)},\tau^{+}(x)<x]}{\partial s}\Big|_{u=0}=\frac{\partial\rho_{+}(s)}{\partial s}[x+\frac{1}{c-\rho_{+}(s)}].$$

The dependencies on $s$ constructed using expressions (61) and (75) are shown in Fig. 12 for $c$=3, $x$=5. These curves practically coincide (except for the turning point of the dependence at $s\approx 0.35$). The dependence on the value of the level $x$ is linear, as in Fig. 2c. From expression (75) the dependence is also determined:

$$\boldsymbol{E}\gamma^{+}(x,u)=\frac{\boldsymbol{E}[\gamma^{+}(x)e^{-u\gamma^{+}(x)},\tau^{+}(x)<x]}{\boldsymbol{E}[e^{-u\gamma^{+}(x)},\tau^{+}(x)<x]}=-\frac{\partial\ln\boldsymbol{E}[e^{-s\tau^{+}(x)-u\gamma^{+}(x)},\tau^{+}(x)<x]}{\partial u}\Big|_{s=0}=\frac{1}{c+u}.$$

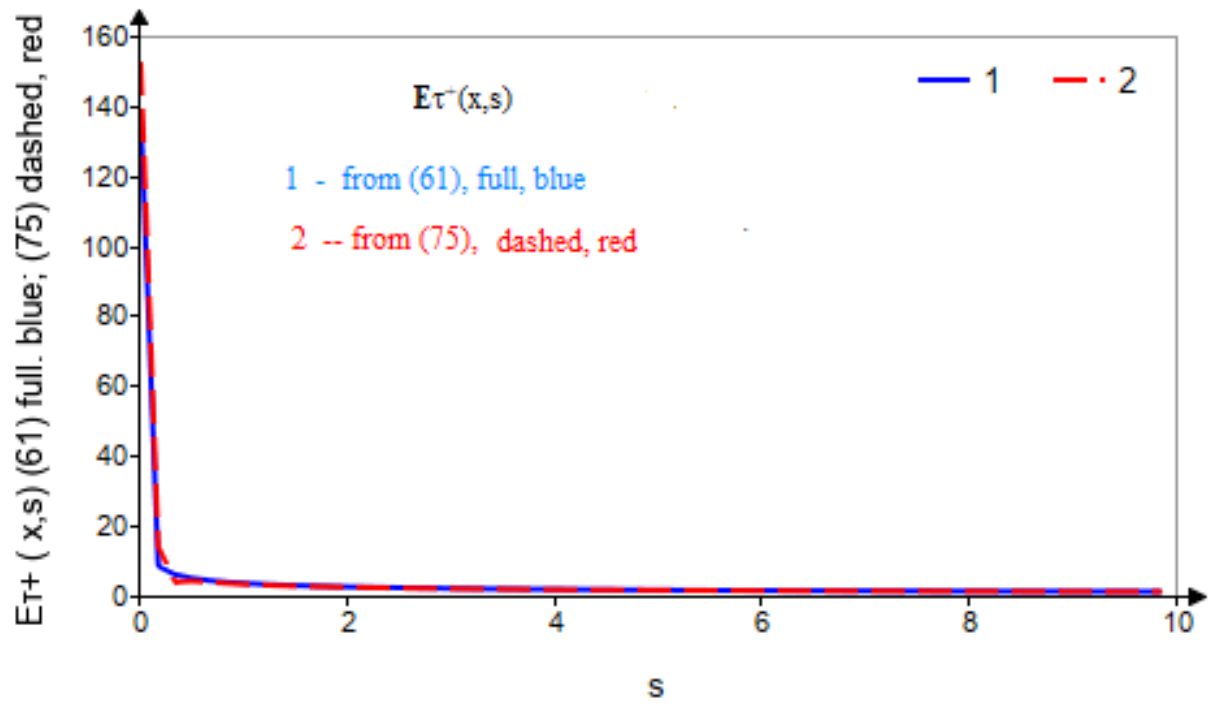

Fig.12. The dependencies of $\boldsymbol{E}\tau^{+}(x,s)$ constructed using expressions (61) (full, blue) and (75) (dashed, red) on $s$ at $c$=3, $x$=5.

The dependence of the duration of the diffusion process $Q_x(t)$ (or $Q_x(\theta_s)$ ) over a fixed level $x$ is also defined. This quantity is specified in expression (23) and below. For the process of Brownian motion, the *MGF* of the quantity $Q_x(\theta_s)$ differs from Eq. (23) and is equal to:

$$D_x(s,\mu)=\boldsymbol{E}e^{-\mu Q_x(\theta_s)}=1-\frac{\rho_+(s+\mu)-\rho_+(s)}{\rho_+(s+\mu)}e^{-\rho_+(s)x},\ x>0, \tag{76}$$

$$\boldsymbol{E}Q_x(\theta_s,\mu)=\frac{\boldsymbol{E}[Q_x(\theta_s)e^{-\mu Q_x(\theta_s)}]}{\boldsymbol{E}[e^{-\mu Q_x(\theta_s)}]}=-\frac{\partial\ln\boldsymbol{E}[e^{-\mu Q_x(\theta_s)}]}{\partial\mu}=\frac{\partial\rho_+(s+\mu)}{\partial\mu}\frac{\rho_+(s)}{(\rho_+(s+\mu))^2}e^{-\rho_+(s)x}/[1-\frac{\rho_+(s+\mu)-\rho_+(s)}{\rho_+(s+\mu)}e^{-\rho_+(s)x}].$$

Fig. 13 shows the behavior of the value $\boldsymbol{E}Q_x(\theta_s,\mu)$ Eq. (76) at $x$=$u$: depending on the parameter $\mu$ for different parameter values: $s$=10, $x$=5 for Fig. 13a), red, ~$10^{-12}$, $s$=1, $x$=1 for Fig. 13a), black, ~$10^{-2}$. Fig. 13b) shows the dependence of the value $\boldsymbol{E}Q_x(\theta_s,\mu)$ on the boundary value $x$ for $s$=5, $\mu$=0.

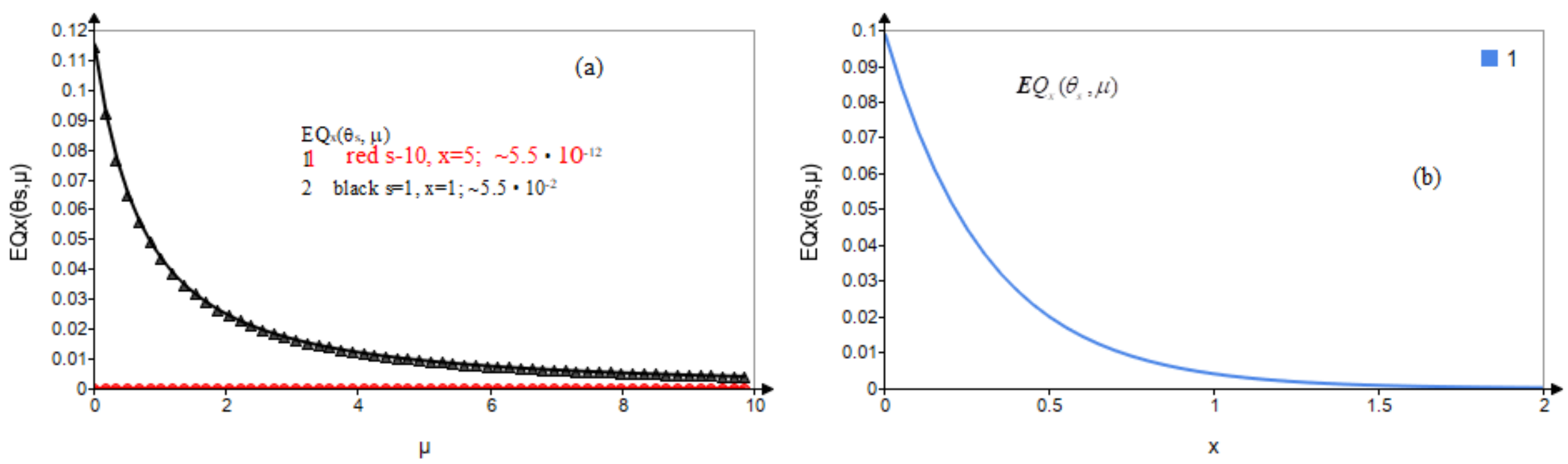

Fig.13a. The behavior of the value $\boldsymbol{E}Q_x(\theta_s,\mu)$ (76): depending on the parameter $\mu$ for $s$=10, $x$=5, red, ~$10^{-12}$; depending on the parameter $\mu$ for $s$=1, $x$=1, black, ~$10^{-2}$.
Fig.13b): The dependence of the value $\boldsymbol{E}Q_x(\theta_s,\mu)$ on the boundary value $x$ for $s$=5, $\mu$=0.

From Fig.13b) it is evident that in this case of the random diffusion process being above a fixed level $x$, the dependence on the value of the boundary value $x$ is nonlinear, an exponential decrease with increasing $x$. In the cases of the time of reaching the level $x$ Fig.2c, Fig.4b, and Eq. (75) this dependence represents a linear increase with increasing $x$. This is due to the fact that expression (76) describes the time spent above the level, and not the time of reaching the level.

### 5.6. Fluctuation magnitudes other than exponential distribution

The results presented in Figure 1 indicate that for an exponential distribution of fluctuation magnitudes, the behavior of the dependence $\rho_{\pm}(s)$ of the process *ζ(t)* jumps can, to some extent, be compared to the results of thermodynamics of trajectories, as noted prior to Figure 1d. Let us consider how non-exponential distributions of the magnitude of fluctuations behave. Thus, Figure 10 illustrates functions $\rho_{\pm}(s)$ for the process of Brownian motion that differ from the results shown in Figure 1.

Let us illustrate the dependence on the distribution of the magnitude of fluctuations, jumps of the process *ζ(t)* (Eq. (2)) using the example of the duration of the process $R(t)$ (Eq. (1)) staying in the lower half-plane (or, equivalently, the duration of the process *ζ(t)* staying above the level *u*), considered in Section 4.6. For simplicity, we consider the level *u*=0, when the general expression (not only for the exponential distribution of the magnitude of fluctuations) takes the form:

$$\tilde{q}(s) = \lim_{u\to 0} \tilde{q}_u(s) = \frac{\lambda}{c}\int_0^{\infty} e^{-z\rho_-(s)}\bar{F}(z)dz\,, \quad \tilde{q}_u(s) = E[e^{-sT'(u)}, \tau^+(u) < \infty],$$

where the distribution of the fluctuation magnitude $\bar{F}(z) = 1 - F(z) = P\{\xi_k > z\}$ can be arbitrary. Figure 6a shows the behavior of the mean value $\boldsymbol{E}[T'(u)e^{-sT'(u)}]_{|u=5}$ Eq. (41) depending on *s* for *u*=5 and at $\bar{F}(z) = e^{-bz}$. The mean value $\boldsymbol{E}[T'(u)e^{-sT'(u)}]_{u=0}$ Eq. (41) depending on *s* for *u*=0 behaves almost the same way. The type and form of the dependence are the same, but the accepted values increase. Here in equation (42) in contrast to expression (41):

$$\boldsymbol{E}T'(s,u) = \frac{\boldsymbol{E}[T'(u)e^{-sT'(u)}, \tau^+(u) < \infty]}{\boldsymbol{E}[e^{-sT'(u)}, \tau^+(u) < \infty]} = -\frac{\partial \ln \boldsymbol{E}[e^{-sT'(u)}, \tau^+(u) < \infty]}{\partial s}.$$

Let us now choose the distribution of the magnitude of fluctuations $\bar{F}(z) = 1 - F(z) = P\{\xi_k > z\}$ in the form of a heavy-tailed distribution, namely a gamma distribution of the form:

$$F(z) = \int_0^z \frac{\beta^{\alpha}}{\Gamma(\alpha)} x^{\alpha-1}e^{-\beta x}dx,\ \ \alpha > 0,\ \ \beta > 0,\ \ \boldsymbol{E}\xi_1 = \frac{\alpha}{\beta}, \quad \boldsymbol{E}e^{i\alpha\xi_1} = [\beta/(\beta - i\alpha)]^{\alpha}.$$

Then:

$$\tilde{q}(s) = \frac{\lambda}{c}\frac{1}{\rho_-(s)}[1 - \frac{1}{(1+\rho_-(s)/\beta)^{\alpha}}],$$

$$\boldsymbol{E}T'(s,u=0) = \frac{\partial \rho_-(s)}{\partial s}[\frac{1}{\rho_-(s)} - \frac{\alpha/\beta}{(1+\rho_-(s)/\beta)((1+\rho_-(s)/\beta)^{\alpha} - 1)}],$$

instead of expression (42). Let us set $\alpha = k = 2$, thus obtaining the Erlang distribution with $k = 2$; *β*=10 for *b*=5. When selecting the parameter *β*, it was assumed that the mean values would be equal to those of the exponential distribution, $1/b = \alpha/\beta$. Then:

$$\boldsymbol{E}T'(s,u=0) = \frac{\partial \rho_-(s)}{\partial s}\frac{(3+\rho_-(s)/\beta)/\beta}{(1+\rho_-(s)/\beta)(2+\rho_-(s)/\beta)}. \qquad (77)$$

It is necessary to find the function $\rho_-(s)$. Recall that when $\alpha = k = 1$ we obtain an exponential distribution. For Lundberg's equation (8) with $k(r) = t^{-1}\ln \boldsymbol{E}e^{r\zeta(t)} = -cr + \lambda(\boldsymbol{E}e^{-r\xi_1} - 1)$, we derive a cubic equation with solutions, one of which is positive and corresponds to the function $\rho_+(s)$, and the other two are negative and coincide.

The approximate solution of this equation, corresponding to the function $-\rho_{_}(s)$, has the form (for the parameter values from expression (28)) $\rho_{_}(s) = 4.168\sqrt{14.095 + 5s} - 14.343$. As a result of the approximation, the equality $\rho_{_}(0) = 0$ is not satisfied.

The behavior of this function is shown in Fig. 14a. We substitute this expression into the relation (77). The mean value $\boldsymbol{ET}'(s,u)$ obtained with this function $\rho_{_}(s)$ is depicted in Fig. 14b.

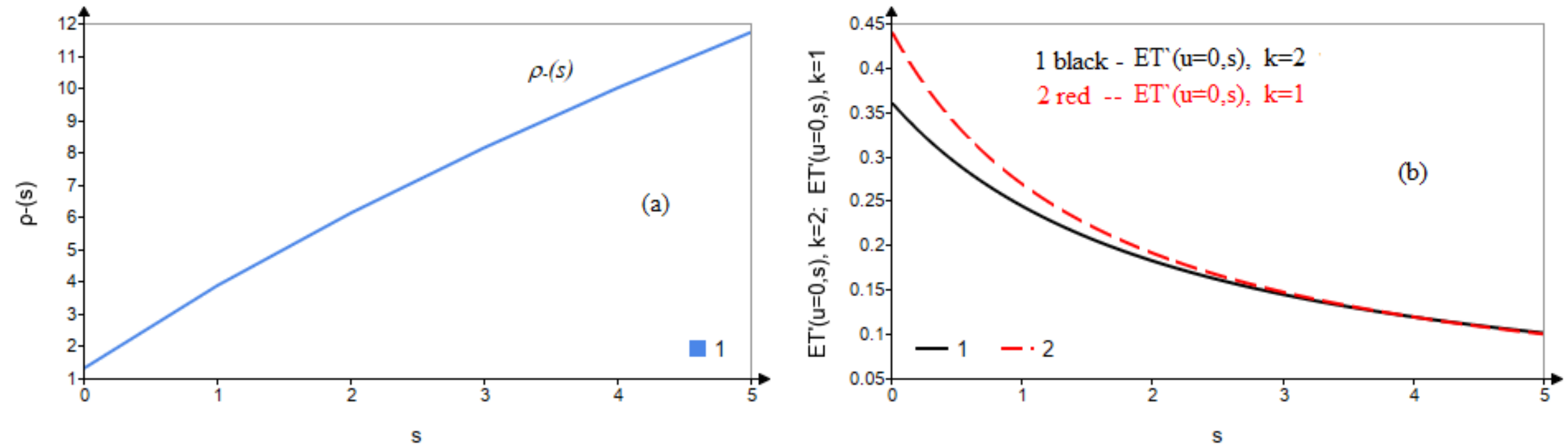

Fig. 14a): the function $\rho_{_}(s)$ for Erlang distribution at $s$>0 with $k$=$\alpha$=2, $\beta$=10; Fig.14b): the mean value $\boldsymbol{ET}'(s,u)$, $u$=0, for Erlang distribution at $s$>0 with $k$=$\alpha$=2, $\beta$=10 (full, black) and with $k$=$\alpha$=1, $b$=5 (dashed, red), exponential distribution.

Let's look at another example of Pareto distribution with survival (tail) function

$$\bar{F}(x) = P\{\xi_k > x\} = \begin{cases} (x_m / x)^\alpha & x > x_m, \\ 1 & x < x_m \end{cases},$$

where $x_m$ is the (necessarily positive) minimum possible value of $\xi_k$, and $\alpha$ is a positive parameter.

The characteristic function is given by:

$$\varphi(t;\alpha,x_m) = \alpha(-ix_m t)^\alpha \Gamma(-\alpha,-ix_m t)$$

where $\Gamma(a,x)$ is the incomplete gamma function. As $\alpha \to \infty$, the distribution approaches $\delta(x - x_m)$ where $\delta$ is the Dirac delta function. The value $x_m$ corresponds to the possible lower limit of the results of jumps, fluctuations of a random variable $\xi_k$. It is assumed that this value is small. With respect to the parameter $\alpha$, make an assumption that this is a large value. In this case, the distribution of the process $\xi_k$ of jumps will look like delta distribution. For $\alpha = n,\ it = r$, large values of $n$, and small values $x = -rx_m$, the expansion of the function $\Gamma(-n,x)$ in a series has the form:

$$\Gamma(-n,x) = x^{-n}[\frac{1}{n} + \frac{x}{1-n} + \frac{x^2}{2(n-2)} + \ldots] + \Gamma(-n) \simeq \frac{x^{-n}}{n}e^{-x} + \Gamma(-n),$$

$$\varphi(it = r;n,x_m) = n(-rx_m)^n \Gamma(-n,-rx_m) \simeq e^{-x} + nx^n\Gamma(-n) \approx e^{-x},$$

since $\lim_{n\to\infty}\Gamma(-n) = 0,\ \lim_{n\to\infty} x^n = 0,\ x < 1$.

The Lundberg equation (8) for the process $\zeta(t)$ Eqs. (2)-(3) takes the form:

$$-cr + \lambda(e^{rx_m} - 1) = s.$$

Considering the smallness of $x_m$, $x = -rx_m$, we expand in a series $e^{-x}$ up to a quadratic term. We obtain a quadratic equation for $r$ of the form:

$$\frac{\lambda}{2} x_m{}^2 r^2 - r(c - \lambda x_m) - s = 0 .$$

The solutions of this equation for $r_+ = \rho_+(s)$ and $r_- = -\rho_-(s)$ are equal to:

$$\rho_+(s) = [\sqrt{(c - \lambda x_m)^2 + 2s\lambda x_m{}^2} + (c - \lambda x_m)] / \lambda x_m{}^2 , \qquad (78)$$

$$\rho_-(s) = [\sqrt{(c - \lambda x_m)^2 + 2s\lambda x_m{}^2} - (c - \lambda x_m)] / \lambda x_m{}^2 .$$

Fig.15 a, b shows the behavior of the functions $\rho_\pm(s)$ for the parameter values: $c$=0.2, $\lambda$=0.819, $x_m$ =0.01. The dependence on $s$ of the average first-passage time to reach level $x$ $\boldsymbol{E}[\tau^+(x), s]$, calculated using expressions (61), (78) is shown in Fig. 15c.

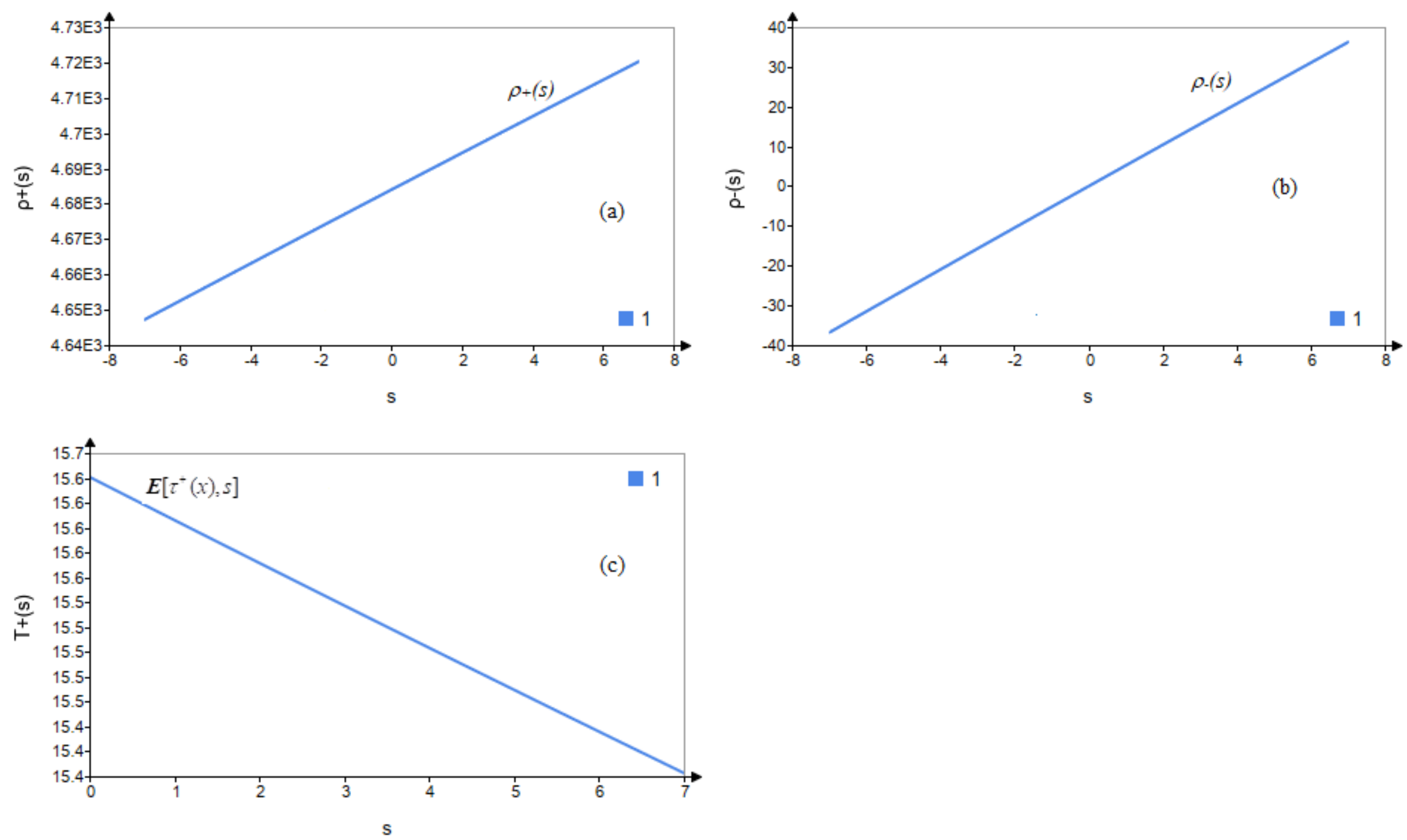

Fig.15 a, b shows the behavior of the functions $\rho_\pm(s)$ for the parameter values: $c$=0.2, $\lambda$=0.819, $x_m$ =0.01, $x$=3; Fig.15c: the dependence on $s$ of the average first-passage time to reach level $x$ $\boldsymbol{E}[\tau^+(x), s]$, $x$=3, calculated using expressions (61), (78).

From Fig. 15 it is evident that for the Pareto distribution of the fluctuation magnitude, the results for $\rho_\pm(s)$ and for $\boldsymbol{E}[\tau^+(x), s]$ differ from the results of Sections 4.2 (for the exponential distribution of the fluctuation magnitude) and 5.3. This may be a consequence of the approximations used in deriving Eq. (78).

Other distributions were also considered, for example, for the Lévy distribution.

## 6. Conclusion

The exploration of boundary functionals beyond first passage times (*FPT*) and extrema of random functions and processes holds significant potential for advancing our understanding and application of statistical physics. While *FPT* and extremal statistics have found an array of applications across various disciplines—ranging from physical systems to complex biological processes—many other boundary functionals remain underutilized despite their direct relevance and physical interpretation. For example, several of the boundary functionals discussed in this paper may find important applications in nuclear reactor theory, where information about boundary functionals such as criticality times, undershoots, overshoots, and dwell times are essential for reactor safety.

*FPT* and extremal statistics have gained substantial traction across various fields, as highlighted by significant contributions in the literature, works [22-58], [57, 89-90]. These studies illustrate their applicability to phase transitions, critical phenomena and stochastic dynamics, yet they also underscore the potential of other boundary functionals that can furnish deeper insights into system behavior under diverse conditions. For instance, boundary functionals can shed light on fluctuation distributions, correlations between physical parameters, and complex interactions in non-equilibrium systems, thus facilitating a more nuanced understanding of such phenomena.

A central aim of this discussion is to broaden the utilization of boundary functionals, advocating for their incorporation into models that predominantly rely on *FPT* and extrema. The tapestry of physical, chemical, and biological systems serves as a fertile ground for the application of these functionals, revealing previously unexplored dimensions of system interactions and responses.

A comprehensive study of boundary functionals could thus pave the way for innovative approaches to longstanding problems in statistical physics and beyond, enriching our toolkit for modeling and understanding complex phenomena. It is crucial for researchers to investigate these alternatives, as they may uncover insights that conventional methods might overlook, yielding a deeper understanding of diverse systems across various fields.

The discussion surrounding boundary functionals in statistical physics has room for expansion, particularly concerning overshooting and undershooting functionals. These functionals, despite their relatively simple mathematical representation, play a crucial role, yet they remain underexplored in the literature, as highlighted by reference [12].

Moreover, there is potential to enrich the analysis of boundary functionals by integrating various methodologies from related fields. For instance, factorization relations and combinatorial techniques may yield new perspectives and enhance the theoretical framework surrounding these functionals.

The application of renewal theory relations has already demonstrated effectiveness in this context, serving as an indicator of the advantages of inter-disciplinary approaches. By harnessing these diverse tools, researchers may be able to address existing gaps and develop a more comprehensive understanding of boundary functionals and their implications in statistical physics. Exploring these connections could lead to novel findings and a richer theoretical landscape, emphasizing the importance of considering a variety of methodologies in the study of complex physical systems.

Extreme values in random processes can serve as the primary functionals, as the distributions of all other boundary functionals (such as *FPT*, sojourn times above a certain level, etc.) are expressed directly through the distributions of these extreme values.

It was noted above that some boundary functionals duplicate each other. This issue should be considered in more detail. Perhaps different definitions will be more convenient in different situations. It is also necessary to consider the correlation links between different boundary functionals: mathematical foundations and physical applications.

Based on the distribution containing *FPT*, it is possible to develop nonequilibrium thermodynamics with a nonequilibrium thermodynamic parameter *FPT* [93–95]. A similar approach can be applied to other boundary functionals, obtaining variants of nonequilibrium thermodynamics associated with them. The text before Fig. 1 also shows the close connection between risk theory and thermodynamics of trajectories. Due to the commonality of the mathematical apparatus, risk theory and boundary functionals can be applied to any physical, biological and other problems.

The article considers a relatively simple case involving an exponential distribution of incoming requests (similar to stochastic storage systems or queueing theory). General expressions are also known [7, 12], but are not provided due to their complexity and the difficulty in obtaining clear, interpretable results from them. In the future, it is necessary to evaluate how much such an approximation affects the final results. In Sections 5.3, 5.5, and 5.6, the problem is considered using a distribution of the magnitude of fluctuations that is different from exponential.

In conclusion, expanding the scope of boundary functionals is critical for elucidating complex systems' behaviors in statistical physics and beyond. By doing so, we may uncover novel dimensions of understanding that traditional methodologies have yet to reveal, ultimately enriching our toolkit for addressing multifaceted scientific questions.